\documentclass[twocolumn]{aastex63}
\usepackage{apjfonts}
\usepackage{amsmath}

\shorttitle{Host Halos of Obscured \& Unobscured QSOs}
\shortauthors{Petter et al.}
\accepted{2023 January 30 to \apj}
\graphicspath{{./}{figures/}}

\begin{document}

\title{Host Dark Matter Halos of \textit{WISE}-selected Obscured 
\& Unobscured Quasars: Evidence for Evolution}

\correspondingauthor{Grayson Petter}
\email{Grayson.C.Petter.GR@dartmouth.edu}

\author[0000-0001-6941-8411]{Grayson C. Petter}
\affil{Department of Physics and Astronomy, Dartmouth College, 6127 Wilder Laboratory, Hanover, NH 03755, USA}

\author[0000-0003-1468-9526]{Ryan C. Hickox}
\affiliation{Department of Physics and Astronomy, Dartmouth College, 6127 Wilder Laboratory, Hanover, NH 03755, USA}

\author[0000-0002-5896-6313]{David M. Alexander}
\affiliation{Centre for Extragalactic Astronomy, Department of Physics, Durham University, South Road, Durham, DH1 3LE, UK}

\author{Adam D. Myers}
\affiliation{Department of Physics and Astronomy, University of Wyoming, Laramie, WY 82071, USA}

\author[0000-0003-4964-4635]{James E. Geach}
\affiliation{Centre for Astrophysics Research, School of Physics, Astronomy \& Mathematics, University of Hertfordshire, Hatfield, AL10 9AB, UK}

\author[0000-0002-8571-9801]{Kelly E. Whalen}
\affil{Department of Physics and Astronomy, Dartmouth College, 6127 Wilder Laboratory, Hanover, NH 03755, USA}

\author[0000-0002-5580-4298]{Carolina P. Andonie}
\affiliation{Centre for Extragalactic Astronomy, Department of Physics, Durham University, South Road, Durham, DH1 3LE, UK}



\begin{abstract}
Obscuration in quasars may arise from steep viewing angles along the dusty torus, or instead may represent a distinct phase of supermassive black hole growth. We test these scenarios by probing the host dark matter halo environments of $\sim 1.4$ million \textit{WISE}-selected obscured and unobscured quasars \added{at $\langle z \rangle = 1.4$} using angular clustering measurements as well as cross-correlation measurements of quasar positions with the gravitational lensing of the cosmic microwave background (CMB). We interpret these signals within a halo occupation distribution (HOD) framework to conclude that obscured systems reside in more massive effective halos ($ \sim 10^{12.9} h^{-1} M_{\odot}$) than their unobscured counterparts ($ \sim 10^{12.6} h^{-1} M_{\odot}$), though we do not detect a difference in the satellite fraction. We find excellent agreement between the clustering and lensing analyses and show that this implies the observed difference is robust to uncertainties in the obscured quasar redshift distribution, highlighting the power of combining angular clustering and weak lensing measurements. This finding appears in tension with models that ascribe obscuration exclusively to orientation of the dusty torus along the line-of-sight, and instead may be consistent with the notion that some obscured quasars are attenuated by galaxy-scale or circumnuclear material during an evolutionary phase.

\end{abstract}



\section{Introduction} \label{sec:intro}

Quasars are the most luminous class of Active Galactic Nuclei (AGNs), manifestations of accretion onto supermassive black holes (SMBHs) at the centers of galaxies. Since their discovery \citep{1963Natur.197.1040S}, AGNs and quasars have been classified into a taxonomy according to their observed multiwavelength properties \citep{2017A&ARv..25....2P}. However, it remains unclear whether or not many of the observed differences between quasar sub-populations reflect intrinsic or cosmologically relevant features of black hole growth.

The class of quasars which are ``obscured'' has been increasingly recognized as important to characterize. These systems often lack the optical and soft X-ray signatures of unobscured quasars, implying that significant columns of dust and gas lie between the sub-pc region surrounding the black hole and our line of sight. A majority of the AGN activity in the Universe is obscured \citep[e.g.,][]{2014ApJ...786..104U, 2015MNRAS.451.1892A, 2015ApJ...802...89B, 2019ApJ...871..240A}, which implies that traditional AGN surveys in the optical or X-rays have potentially led to a biased picture of SMBH growth. Some of the most pressing questions in extragalactic astrophysics require a complete census of AGN activity extending to even the most heavily obscured systems, including understanding the physical structure of AGN systems, probing the cosmic history of black hole growth, and uncovering the role that AGNs play in galactic evolution \citep{2018ARA&A..56..625H}. Crucial to these endeavors is understanding the nature of the obscuring material and whether obscured quasars represent intrinsically different systems to their unobscured counterparts.

The canonical view of the AGN structure is termed the ``unified model'' \citep{1993ARA&A..31..473A, 1995PASP..107..803U, 2000ApJ...545...63E, 2015ARA&A..53..365N, 2017NatAs...1..679R}, which posits that most AGN systems consist of a similar axisymmetric structure of material surrounding the SMBH, but appear to differ because of our chance line-of-sight toward this structure. This model implies that a quasar will appear obscured when our viewing angle happens to nearly coincide with the plane of the dusty ``torus''. In this view, obscured and unobscured quasars represent intrinsically similar objects and are thus expected to occupy similar environments, even when accounting for potential relationships between accretion rate and torus covering factor \citep{2020ApJ...888...71W}. 

Alternatively, evolutionary models suggest that AGNs vary in their observed properties over the course of their lifetimes through interaction with their broader environments. In this scenario, obscured quasar activity could represent a specific evolutionary stage of SMBH growth. One such model posits that rapid star formation and SMBH growth are triggered during major galaxy mergers which funnel gas to the nuclear region of the merger remnant. This process is expected to generate quasar activity obscured by galactic or circumnuclear star-forming gas, before this activity heats and expels the gas to unveil an unobscured quasar \citep[e.g.,][]{1988ApJ...325...74S, 2005ApJ...630..705H, 2006ApJS..163....1H, 2008ApJS..175..356H, 2012NewAR..56...93A, 2018ARA&A..56..625H}. A number of recent studies have found evidence for AGN obscuration taking place on galactic scales or correlated with star formation activity
\citep[e.g., ][]{2015ApJ...802...50C, 2017MNRAS.464.4545B, 2017MNRAS.468.1273R, 2019A&A...623A.172C, 2021ApJ...918...22L, 2021ApJ...914...83Y, andonie22, 2022arXiv220603508G,  2022ApJ...925..203J}, which may be connected to this evolutionary scenario.

A natural test of these models lies in measuring how quasars populate the large-scale structure (LSS) of the Universe, or the manner in which they occupy dark matter halos. This is because the properties of a quasar-hosting halo could not feasibly be connected with the obscuring torus' orientation along our particular line-of-sight, while a connection between halo properties and galaxy evolution is expected. Thus, this work aims to probe the nature of infrared-selected obscured and unobscured quasars by estimating the host dark matter halo properties of each class. 

The connection between halo properties and obscuration is contested. Some studies find that obscured systems occupy more massive halos \citep{2011ApJ...731..117H, 2012A&A...537A.131E, 2014ApJ...789...44D, 2014MNRAS.442.3443D, 2015MNRAS.446.3492D, 2016MNRAS.456..924D, 2017MNRAS.469.4630D, 2018ApJ...858..110P}, some find the opposite \citep{2010ApJ...716L.209C, 2014ApJ...796....4A}, and others find no trend \citep{2009ApJ...701.1484C, 2009A&A...494...33G, 2009A&A...500..749E, 2012MNRAS.420..514M, 2013ApJ...776L..41G, 2016ApJ...821...55M, 2016ApJ...832..111J, 2018MNRAS.481.3063K, 2018MNRAS.474.1773K}. These studies, though, vary in their selection, statistical power, analysis technique, obscuration definition, and sample distributions across luminosity and redshift.

In this work, we utilize the largest sample of mid-infrared-selected quasars to date in order to put precise constraints on the host halo properties of obscured and unobscured quasars \added{at $\langle z \rangle = 1.4$}. We probe these properties first by measuring angular autocorrelation functions, and interpreting these signals in a halo occupation distribution (HOD) framework. We find that obscured quasars occupy significantly more massive halos than unobscured quasars on average, but we do not detect a difference in the fraction that are satellites. Next, we test the halo properties with an entirely independent method, the cross-correlation of quasar positions with \textit{Planck}'s map of the gravitational lensing of the CMB. We interpret the lensing signals with a linearly-biased model, and again find that obscured quasars occupy significantly more massive halos. The implied effective halo masses from the clustering and lensing analyses are in excellent agreement, which we show implies our results are robust against uncertainties in the obscured quasar redshift distribution. We interpret these results as favoring an evolutionary explanation for the obscuration of at least some quasars.


Throughout this work, we adopt a ``Planck 2018'' CMB+BAO $\Lambda$-CDM concordance cosmology \citep{2020A&A...641A...6P}, with $h = H_0/100 \ \mathrm{km \ s}^{-1} \mathrm{Mpc}^{-1} = 0.6766$, $\Omega_{m} = 0.3111$, $\Omega_{\Lambda} = 0.6888$, $\sigma_{8} = 0.8102$, and $n_{s} = 0.9665$.

\section{Data} \label{sec:data}
\subsection{WISE Quasar Sample}
\label{sec:sample}
Though heavily obscured quasars can be challenging to distinguish from normal galaxies in the optical waveband, they can be recovered simply via their red mid-infrared colors \citep{2004ApJS..154..166L, 2005ApJ...631..163S, 2012ApJ...753...30S, 2012ApJ...748..142D, 2013ApJ...772...26A} which trace the reprocessed emission from the dusty torus. The Wide-field Infrared Survey Explorer \citep[\textit{WISE;}][]{2010AJ....140.1868W} has thus revolutionized obscured quasar selection by mapping the entire sky in four mid-infrared bands centered at (named) 3.6 (W1), 4.5 (W2), 12 (W3), and 22\,$\mu$m (W4). These data currently provide the only diagnostic capable of producing highly reliable samples of millions of both obscured and unobscured quasars over the entire extragalactic sky. 

We therefore elect to use the \citet{2018ApJS..234...23A} 90\% reliable (R90) criterion on \textit{WISE} photometry to select our parent sample of quasars. Rather than adopting the publicly released catalog associated with \citet{2018ApJS..234...23A} which used the criterion to select candidates from the AllWISE catalog \citep{2014yCat.2328....0C}, we apply the criterion to the newer CatWISE 2020 \citep{2020ApJS..247...69E, https://doi.org/10.26131/irsa551, 2021ApJS..253....8M} data release. This catalog is generated from \textit{WISE} observations taken between 2010 and 2018, incorporating six times more exposures in the W1 and W2 channels than were used in generating the AllWISE catalog. This deeper imaging enables more precise measurements of photometric colors and therefore more reliable selection of quasars. Crucially, the deeper imaging will also enable a uniformly complete selection across the sky to fainter fluxes, greatly simplifying the reconstruction of the selection function necessary to perform a clustering measurement. We thus query the IRSA\footnote{https://irsa.ipac.caltech.edu} database and select all objects in the CatWISE catalog satisfying the R90 criterion, adopting \textbf{mpro} Vega magnitudes. We also apply a magnitude limit in the W2 channel such that the selection is $> 99\%$ complete across the entire sky, and a bright-end cut to exclude infrared stars:

\begin{equation}
    \left\{ \begin{aligned}
    &\mathrm{W}1 - \mathrm{W}2 > 0.65 \times \mathrm{exp}[0.153 \times (\mathrm{W}2 - 13.86)^{2}], \\
    &9 < \mathrm{W}2 < 16.5
    \end{aligned} \right.
    \label{eq:select}
\end{equation}

As the \citet{2018ApJS..234...23A} criterion was calibrated in the extragalactic Bo{\"o}tes field where Galactic objects are relatively rare, we must apply masks to remove regions which likely suffer from higher contamination rates. For this, we follow the procedure described by \citet{2018ApJS..234...23A}, though we create multi-order coverage (MOC) maps \citep{2014ivoa.spec.0602F} to accomplish this rather than remove sources geometrically. In particular, we mask regions within 10$^{\circ}$ of the Galactic plane and within 30$^{\circ}$ of the Galactic center. We also mask regions occupied by Galactic Planetary Nebulae \citep{1992secg.book.....A}, $\mathrm{H}_{\mathrm{II}}$ regions \citep{2014ApJS..212....1A}, star-forming regions \citep{1962ApJS....7....1L, 1965ApJS...12..163L}, and resolved nearby galaxies in the Catalog and Atlas of the Local Volume Galaxies \citep[LVG; ][]{2013AJ....145..101K} or the 2MASS extended source catalog \citep[XSC; ][]{2006AJ....131.1163S}. We refer the reader to \citet{2018ApJS..234...23A} for more detail in this masking procedure. We will refer to this mask as the ``contamination mask'' throughout the remainder of this work.

One of the advantages of the CatWISE 2020 catalog over previous \textit{WISE} releases is the order of magnitude improvement in detecting proper motions owing to the longer time baseline. This information allows the removal of stellar or solar system object contaminants from our quasar candidate sample. We thus remove sources with measured motions $ > 0.25 \arcsec/\mathrm{yr}$, which are able to be detected at $> 5 \sigma$ across the sky down to our flux limit. This cut removes $1.5 \%$ of objects from the masked catalog.

\subsubsection{Optical Data: Binning by Obscuration} \label{sec:parsample}
The distribution of optical to mid-infrared colors of infrared-selected quasars appears bimodal, as rest-frame UV-optical emission is easily extincted by dust, while near-infrared emission is less so. Therefore, a simple color cut can be used to classify obscured systems. Quasars with optical-infrared colors $r - $W2 $> 3.1$ [AB] typically show X-ray absorption corresponding to absorbing column densities of $N_{\mathrm{H}} > 10^{22} $ \citep[][]{2007ApJ...671.1365H}. In order to classify our sources as "obscured" or "unobscured," we sought an optical survey deep enough to detect most of the quasars and as wide as possible to maintain a large sample size. We therefore utilize $r$-band optical data from the ninth release (DR9) of the Dark Energy Spectroscopic Instrument Legacy Imaging Survey \citep[DESI-LS; ][]{2019AJ....157..168D}. This survey covers $\sim 20,000 \ \mathrm{deg}^2$ of extragalactic sky to $r_{\mathrm{AB}} \sim 24$, and the unprecedented combination of depth and area ensures that we are able to estimate the degree of obscuration for a uniform sample consisting of greater than one million quasars for the first time. We match our parent sample of \textit{WISE} quasars to the DESI-LS catalog with a matching radius of $2 \arcsec$ \citep{2013ApJ...772...26A} using NOIRLab's Astro Data Lab\footnote{https://datalab.noirlab.edu} tools. The optical photometry is corrected for Galactic reddening using the map of \citet{1998ApJ...500..525S}. We retain the $\sim 8\%$ of \textit{WISE} quasar candidates within the DESI-LS footprint lacking optical counterparts and assign lower limits to their $r-W2$ color indices based on the local $r$-band imaging depth. To estimate the color distribution of sources undetected in DESI-LS, we match the non-detections to the Hyper Suprime-Cam Subaru Strategic Program (HSC-SSP) ``DEEP'' catalog \citep{2018PASJ...70S...8A} in the areas where the surveys overlap. 96$\%$ of the DESI-LS non-detections are detected in HSC-DEEP, implying a negligible contamination rate of spurious WISE sources in our sample.

\begin{figure}
    \centering
    \includegraphics[width=0.45\textwidth]{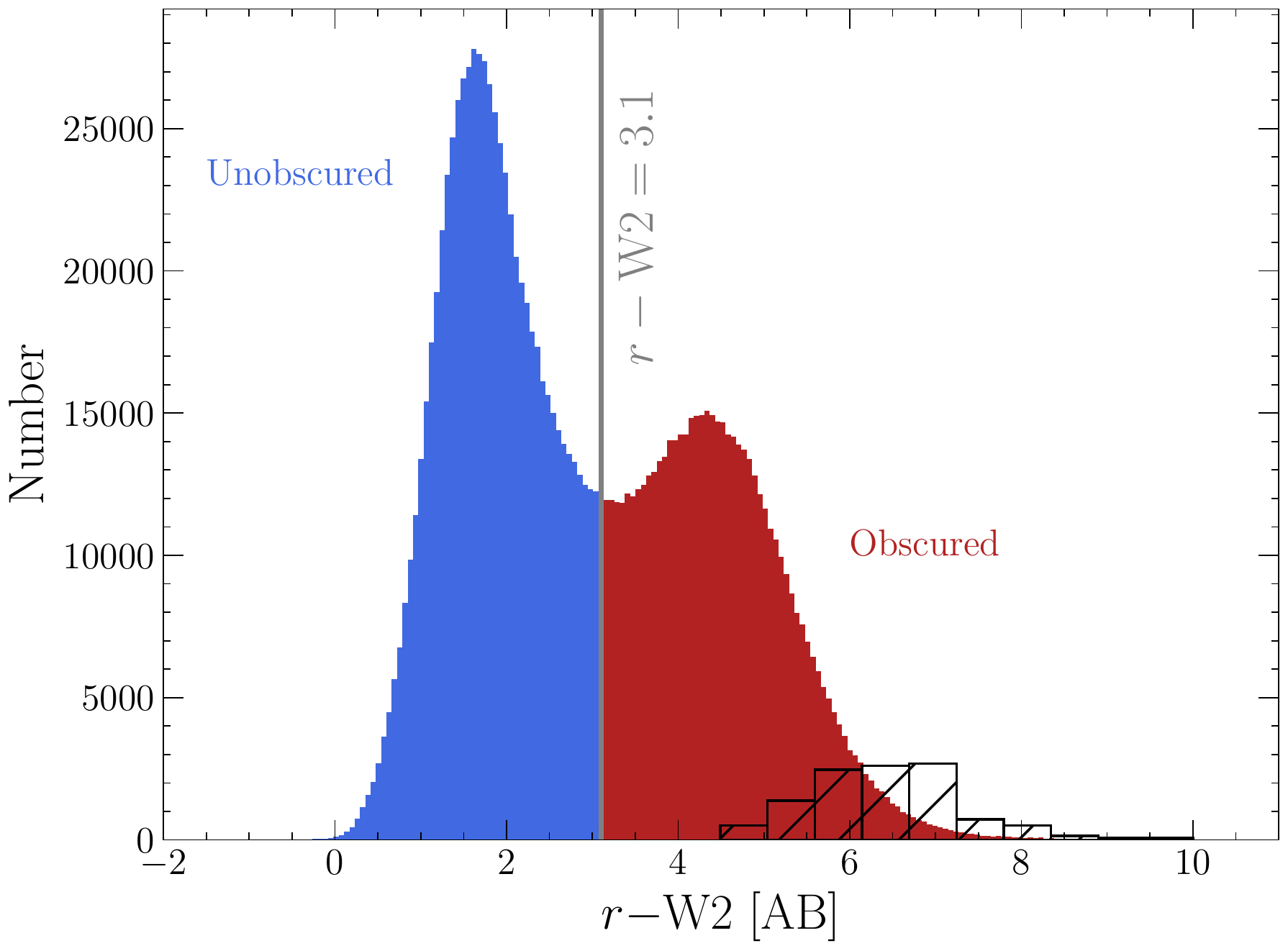}
    \caption{The optical-infrared ($r-$W2) color distribution of 1.4 million \textit{WISE}-selected quasar candidates. The distribution is clearly bimodal, reflecting that infrared quasar selection recovers both obscured and unobscured populations. We classify quasars with $r-$W2 > 3.1 [AB] as obscured, displaying them in red, and show unobscured quasars in blue. We note that all masking described in $\S$\ref{sec:sample} and contaminant removal ($\S$\ref{sec:purity}) has been applied before producing this distribution. We also show the color distribution of the $8\%$ of sources not detected in DESI-LS using black hatched boxes, obtained by matching against the HSC-SSP DEEP catalog.}
    \label{fig:colors}
\end{figure}

The resulting optical-infrared color distribution of our quasar sample is shown in Figure \ref{fig:colors}. It is clear that \textit{WISE} mid-infrared selection reveals a population with a bimodal optical-infrared color distribution, and thus uncovers both obscured and unobscured sources in roughly equal proportion. We adopt the criterion of \citet{2007ApJ...671.1365H} ($r - \mathrm{W}2$ [AB] > 3.1) to separate the sample into obscured and unobscured subsets. This results in 52$\%$ of the sample being classified as unobscured, and $48\%$ as obscured.

\subsection{Sample Purity}
\label{sec:purity}

In order to robustly interpret clustering measurements of quasar candidates, it is important to optimize and quantify the purity of the quasar selection. Mid-infrared quasar selection can be contaminated by low-redshift star-forming galaxies (SFGs) as well as by luminous high-redshift galaxies. \citet{2021ApJ...922..179B} showed that the most significant source of sample contamination in the \citet{2018ApJS..234...23A} R90 catalog is from $z = 0.2 - 0.3$ galaxies with high specific star formation rates. \citet{2012ApJ...754..120M} developed a criterion to discriminate between low redshift star-forming galaxies and AGN using a near-infrared to mid-infrared color, $K - [4.5]$. In this work, we show that an alternative three-band diagnostic can effectively isolate low-redshift star-forming contaminants, as these objects are redder at $r - W2$ versus $z - W2$ than quasars. This was discovered by observing that optically-bright candidates appear bimodal in this color space, with objects redder at $r - W2$ compared to $z - W2$ having photometric redshifts $z < 0.7$ \citep{2021ApJ...922..179B} and appearing resolved in DESI-LS imaging.

\begin{figure}
    \centering
    \includegraphics[width=0.45\textwidth]{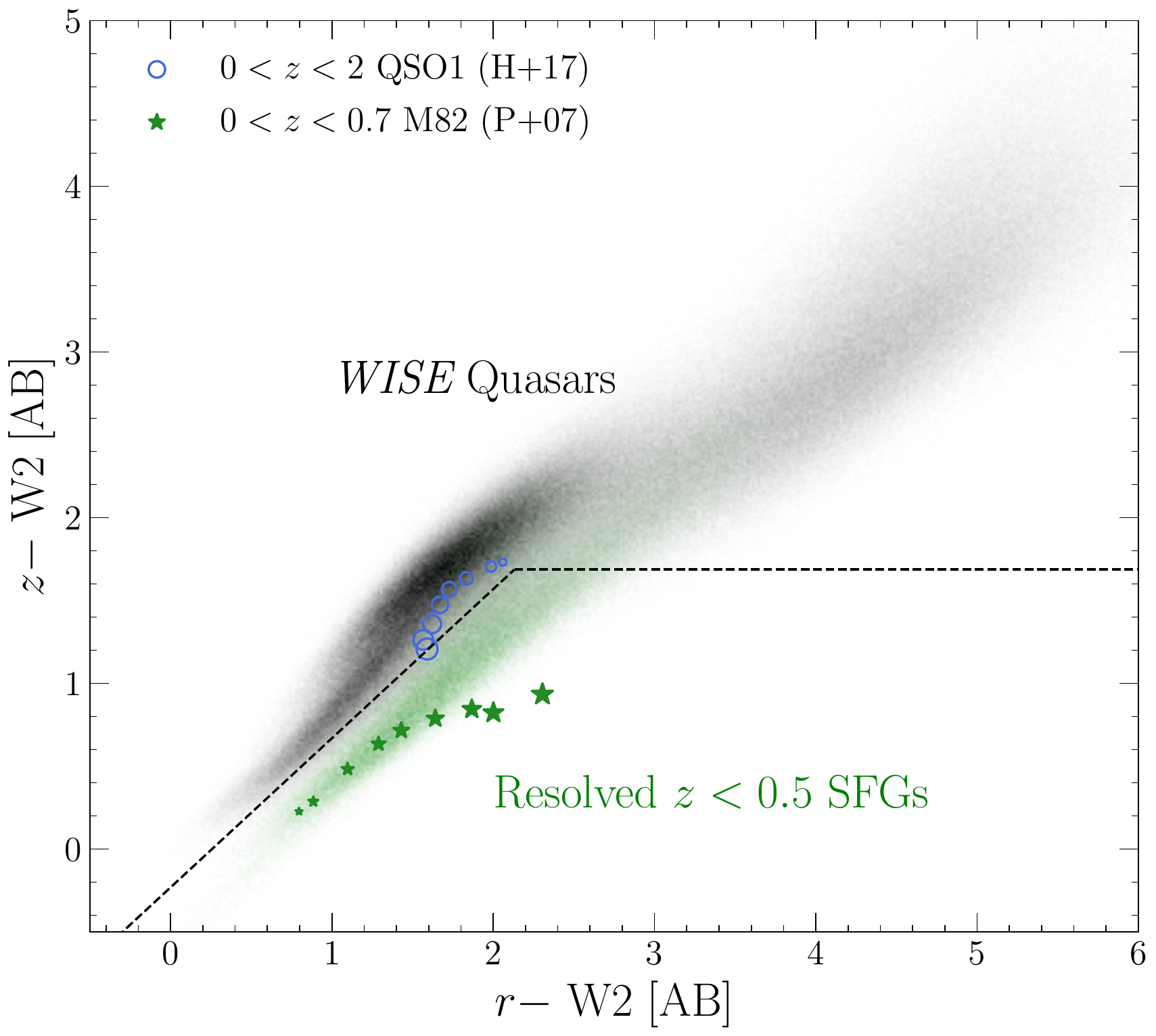}
    \caption{The distribution of \citet{2018ApJS..234...23A} R90 infrared quasar candidates in an optical ($r$), near-infrared ($z$), mid-infrared (W2) color space. We show the density of candidates in greyscale, and in green we show the density of highly resolved sources in the optical band representing low-redshift galaxy contaminants. Unobscured candidates ($r$-W2 < 3.1) appear bimodal roughly along a line of constant $r-z$, which differentiates low-redshift SFG contaminants from quasars. Template colors of Type-1 quasars \citep{2017ApJ...849...53H} at $0 < z < 2$ and the starburst M82 at $0 < z < 0.7$ \citep{2007ApJ...663...81P} are shown with blue open circles and green stars, respectively, and increasing marker size represents linear steps towards increasing redshift. We show our criterion to remove low-$z$ star-forming galaxy contaminants with a black dashed line.}
    \label{fig:contam}
\end{figure}

We display our candidates in this color space in Figure \ref{fig:contam}. Using a separate color, we show candidates which are highly resolved in the optical imaging, defined as being significantly better fit with a round exponential profile (REX) than as a point source ($\chi^2_{\mathrm{REX}} -\chi^2_{\mathrm{PSF}} > 1000$). We also show predicted colors of optical quasars using the Type-1 template of \citet{2017ApJ...849...53H}, and of the prototypical starburst galaxy M82 from \citet{2007ApJ...663...81P}. It is clear that the optically-bright end of the R90 quasar selection criterion includes low-redshift galaxy-dominated contaminants. We use a simple intersection of two lines to excise sources in the SFG region of color space, which removes $12\%$ of our full sample. We note that the quasar color distribution shown in Figure \ref{fig:colors} was produced after culling these galaxy contaminants.

We briefly consider the reliability of this new catalog by matching it with the photometric redshift catalog produced in the Bo{\"o}tes field \citep{2021A&A...648A...4D}. This catalog contains AGN diagnostic flags for every source denoting whether it appears in the `Milliquas' complilation of spectroscopically confirmed quasars from the literature \citep{2019arXiv191205614F}, exhibits mid-infrared \textit{Spitzer}-IRAC colors indicative of AGN activity \citep{2012ApJ...748..142D}, or coincides with a luminous X-ray source \citep{2005ApJS..161....9K}. 95\% of our sample sources pass at least one of these criteria, while only 75\% of sources in the \citet{2018ApJS..234...23A} R90 catalog do. We attribute this reliability increase to the more precise CatWISE photometric colors as well as our removal of galaxy contaminants. Our catalog is thus expected to be highly reliable and any systematics introduced into our clustering measurements from non-quasar contamination will be subdominant.

\subsection{Redshift Information}

Interpreting the angular clustering or CMB lensing signal for a sample of quasars requires knowledge of the sample's redshift distribution. Targeted quasar redshift surveys are currently limited to optically-bright relatively unobscured quasars \citep[e.g.,][]{2020ApJS..250....8L, 2022arXiv220808517A, 2022arXiv220808511C}, and individual blind redshift surveys lack either the breadth or depth to constrain the redshift distribution of our full sample. In order to obtain a nearly-complete and statistically representative estimate of the redshifts of our sample quasars, we therefore look to the well-characterized Bo{\"o}tes and COSMOS \citep{2007ApJS..172....1S} fields. 

We select the portion of our sample falling within the footprint of the AGN and Galaxy Evolution Survey \citep[AGES,][]{2012ApJS..200....8K} in Bo{\"o}tes or the zCOSMOS \citep{2007ApJS..172...70L} footprint in the COSMOS field. We then performed a literature search for photometric and spectroscopic redshifts of the objects in these fields by matching them against various publicly-available redshift catalogs with a $2\arcsec$ search radius. We prioritize spectroscopic over photometric measurements, and adopt the spectroscopic estimate with the highest quality flag when available. We also reject photometric redshift estimates of $z > 3$ as unrealistic and likely the result of poor fits. In Bo{\"o}tes, we take photometric redshifts from \citet{2014ApJ...790...54C}, and then \citet{2021A&A...648A...4D}, with the latter taking priority. In COSMOS, we adopt photometric redshifts from \citet{2017A&A...602A...3D} and then \citet{2016ApJ...817...34M}, which in turn depend on analyses from \citet{2009ApJ...690.1250S, 2011ApJ...742...61S}, \citet{2009ApJ...690.1236I}, and \citet{2016ApJS..224...24L}. The surveys providing the spectroscopic redshifts include AGES \citep{2012ApJS..200....8K}, SDSS DR16 \citep{2020ApJS..249....3A}, zCOSMOS \citep{2007ApJS..172....1S}, PRIMUS \citep{2011ApJ...741....8C, 2013ApJ...767..118C}, IDEOS \citep{2016MNRAS.455.1796H}, FMOS-COSMOS \citep{2019ApJS..241...10K},
DEIMOS10K \citep{2018ApJ...858...77H}, and the studies of \citet{2006ApJ...644..100P}, \citet{2009ApJ...696.1195T},
\citet{2012ApJ...744..150B}, \citet{2012ApJ...761..140C}, \citet{2013ApJS..208...24L},
\citet{2013ApJ...772...26A},  \citet{2015A&A...575A..40C},
\citet{2015ApJ...806L..35K}, \citet{2015ApJ...808..161O}, and \citet{2018ApJS..239...22S}. Overall, we find that 100$\%$ and $97\%$ of unobscured and obscured quasars of the 899 total sources in these fields have a redshift measurement, respectively. We thus treat the redshift distributions as representative of their respective samples.

\begin{figure}
    \centering
    \includegraphics[width=0.45\textwidth]{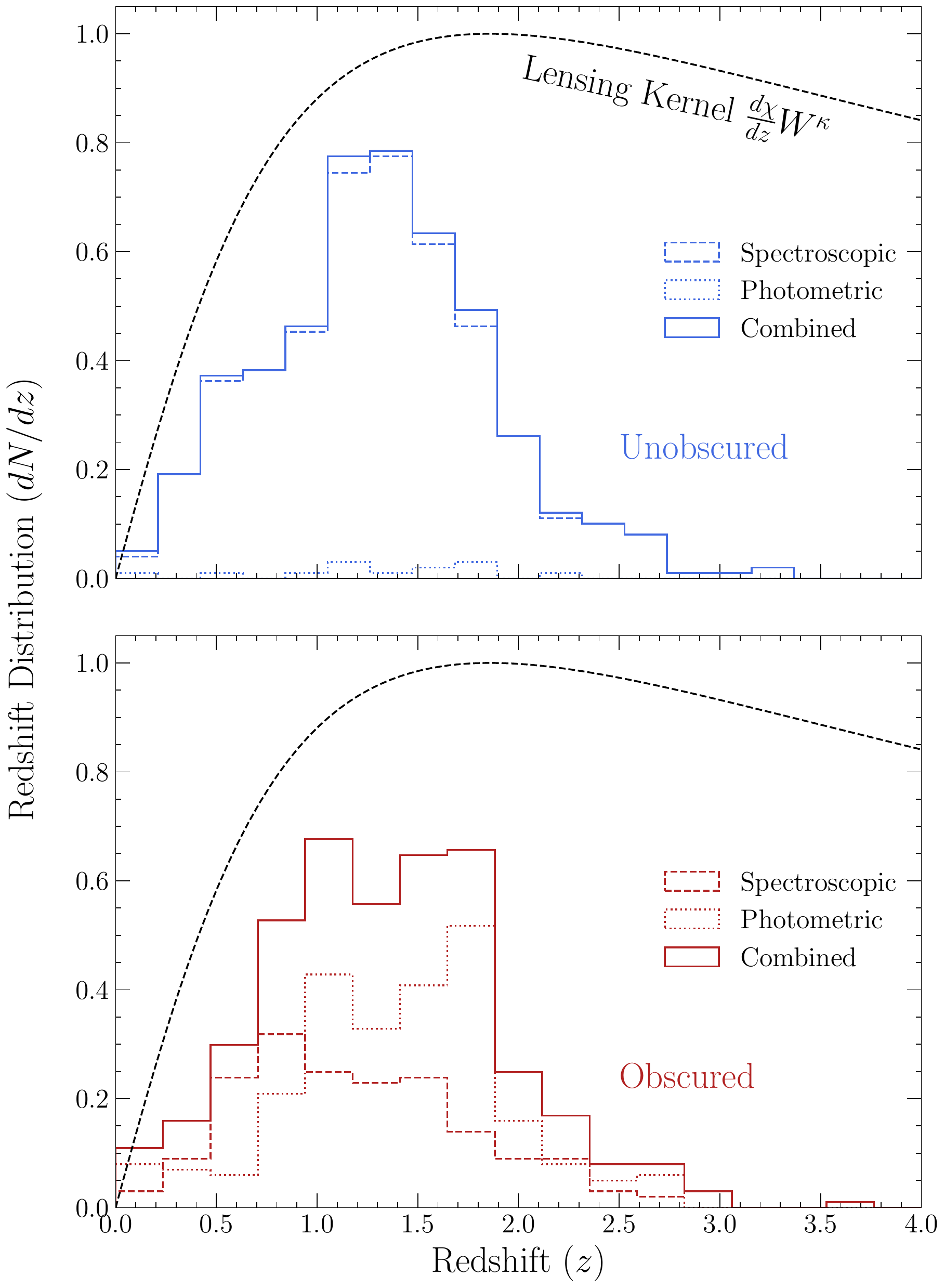}
    \caption{The redshift distributions $(dN/dz)$ of the unobscured (top panel in blue) and obscured (bottom panel in red) quasar samples obtained by cross-matching to surveys in the Bo{\"o}tes and COSMOS fields. The samples \replaced{appear}{are} similarly distributed in redshift \added{(a KS test p-value of $p=0.15$ fails to reject the null hypothesis that the samples are drawn from the same distribution)}, implying that we can appropriately compare their clustering properties. Redshift distribution information stemming from spectroscopic redshifts is shown with a dashed line, a dotted line shows information from photometric redshifts, and a solid line shows the combinations. The unobscured distribution is derived mostly from spectroscopic redshifts, while we rely on photometric redshifts for half of the obscured population. We also display the shape of the CMB lensing kernel (Eq.\ \ref{eq:lenskern}) with a dashed black line, showing that the LSS surrounding our samples should efficiently lens CMB photons.}
    \label{fig:zdist}
\end{figure}

We show the resulting redshift distributions of the obscured and unobscured samples in Figure \ref{fig:zdist}. Incorporating both photometric and spectroscopic redshifts, the distributions appear similar between the samples, implying their clustering properties can be appropriately compared. \added{We verify this by measuring a p-value of $p=0.15$ in a KS test, failing to reject the null hypothesis that the two samples are drawn from the same distribution}. However, only photometric redshifts are available for approximately half of the obscured sample. Therefore, we will test the robustness of our clustering results against possible systematics in the obscured population's redshift distribution in $\S$\ref{sec:discussion}.

\subsection{Planck Lensing Convergence Map}
The CMB radiation from $z \approx 1090$ has been gravitationally lensed by the intervening large-scale structure (LSS), and thus encodes information about the dark matter halos that host galaxies across cosmic time. Therefore, measuring the cross-correlation between the lensing convergence and the angular overdensity of quasars independently constrains the sample's host halo properties. In this work, we utilize the 2018 release of the CMB lensing convergence map \citep{2020A&A...641A...8P} provided by data from the \textit{Planck} satellite to probe the halo properties of quasars as a function of obscuration. We adopt the map generated using the minimum-variance (MV) estimator, which used both temperature and polarization data to reconstruct the lensing convergence. We also make use of simulated noise maps provided with the data release to estimate uncertainties. To generate the lensing and noise maps, we transform the provided $\kappa_{\ell m}$ coefficients at $\ell < 2048$ into NSIDE=1024 resolution {\tt HEALPix} maps.

\subsection{Ancillary Photometry}
\label{sec:lums}
\begin{figure}
    \centering
    \includegraphics[width=0.45\textwidth]{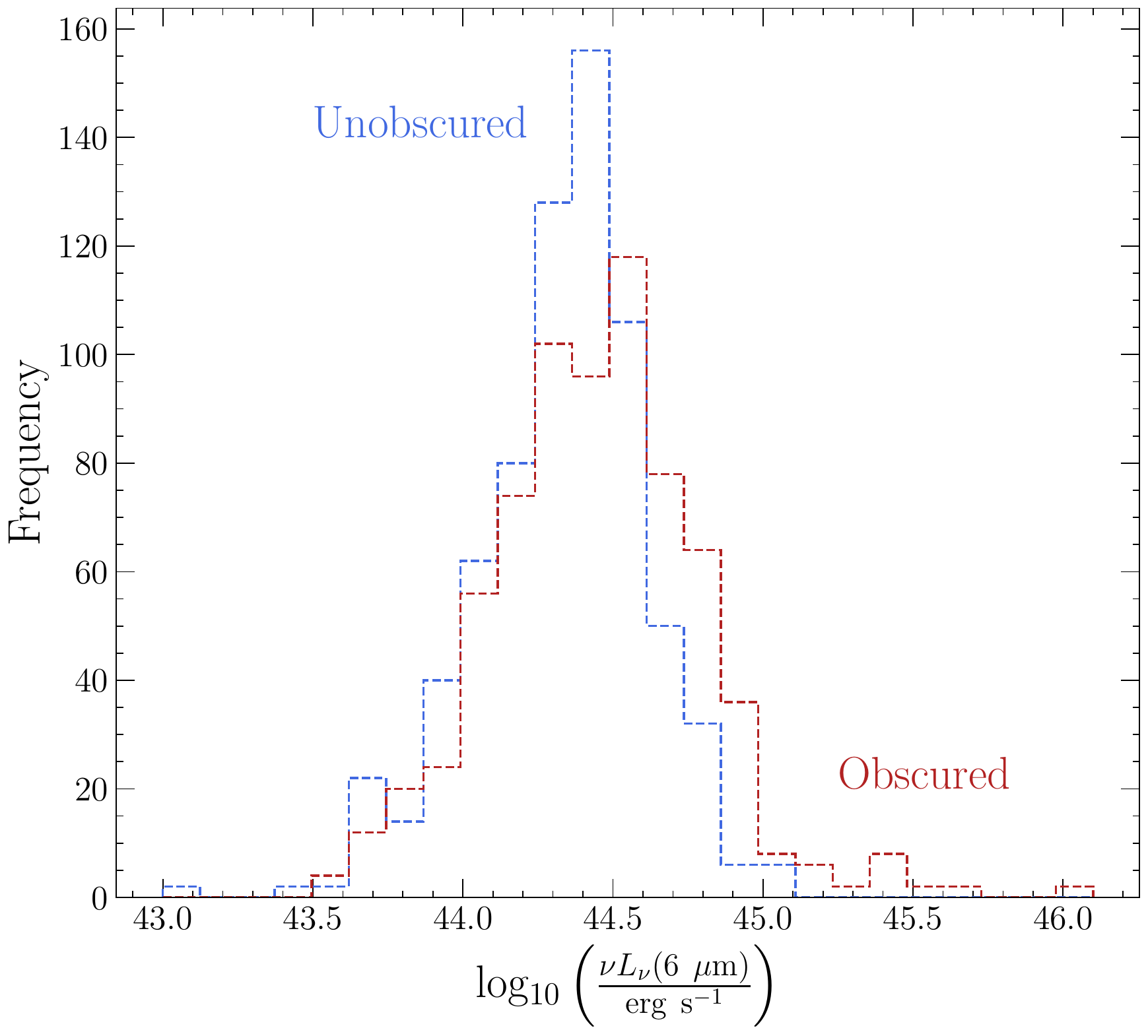}
    \caption{\added{The $6 \ \mu$m luminosity distributions for the obscured and unobscured quasar samples computed using SDWFS mid-infrared photometry. The samples are similarly distributed in luminosity, with a modest tail towards higher luminosities in the obscured sample. We utilize these distributions along with the clustering measurements in this work to study the occupation statistics of each subset in $\S$\ref{sec:nto}.}}
    \label{fig:lum_dists}
\end{figure}

Finally, we estimate the bolometric luminosity distributions of obscured and unobscured quasars to compute space densities and occupation statistics in $\S$\ref{sec:nto}. We use the rest-frame $6 \ \mu$m band as a bolometric luminosity tracer which is minimally affected by obscuration. As $90\%$ our sample falls within the \citet{2012ApJ...748..142D} AGN wedge, we suggest this wavelength is also primarily tracing the torus luminosity and is uncontaminated by star formation activity. For checking the location of these sources in \textit{Spitzer} color-color space, we utilize mid-infrared \textit{Spitzer} photometry for our sources from the \textit{Spitzer} Deep, Wide-field Survey \citep[SDWFS;][]{2009ApJ...701..428A} as collated by the \textit{Herschel} Extragalactic Legacy Project \citep[HELP, ][]{2019MNRAS.490..634S, 2021MNRAS.507..129S}. 
\added{We display the $6 \ \mu$m luminosity distributions for obscured and unobscured quasars in Figure \ref{fig:lum_dists}}.

\section{Measurements}

\subsection{Masking}
\label{sec:masking}
\added{Clustering measurements are inherently sensitive to systematics arising from a non-uniform selection function across the sky. Therefore,} in addition to the contamination mask generated following \citet{2018ApJS..234...23A}, we create several more masks to excise regions likely to introduce systematics into our measurements. As we permit optically-undetected quasars in our sample, we must carefully characterize the imaging footprint of DESI-LS to avoid classifying quasars as obscured simply because they lack deep optical imaging. Fortunately, DESI-LS DR9 provides random catalogs \citep{2022arXiv220808518M}\footnote{https://portal.nersc.gov/cfs/cosmo/data/legacysurvey/dr9/randoms/} populated within the imaging footprint, with an associated local depth for each random point and each band. We take the median of the $r$-band depth in {\tt HEALPix} cells, and mask cells where the 5-$\sigma$ depth is $r < 23$. Next, we create a ``reddening mask'', excising regions with a high degree of Galactic reddening, $E(B-V) > 0.1$ \citep{1998ApJ...500..525S}. All of these masks are generated with {\tt HEALPix} at NSIDE=1024 resolution.

We must also remove regions with severe \textit{WISE} imaging artifacts, such as diffraction spikes, saturated pixels, latents and ghosts. However, these regions are often localized to scales of a few arcseconds, so masking the parent NSIDE=1024 pixel (corresponding to a 3.4$\arcmin$ resolution) of every flagged region would cause us to remove an unacceptably large fraction of the sky from our analysis. Instead, we use the UnWISE bitmasks \citep{2019PASP..131l4504M} to remove sources in the data and random catalogs overlapping with flagged pixels. This process, as well as the contamination masking procedure ($\S$ \ref{sec:sample}) effectively masks a fraction of each parent pixel, which we must correct for when generating the quasar overdensity map to cross-correlate with the CMB lensing map. Thus, we compute the fractional area lost within each NSIDE=1024 pixel, naming this the ``lost-area'' mask. We further apply a boolean mask to remove pixels in which the lost area is greater than 20\% \citep[e.g., ][]{2020JCAP...05..047K}.

\replaced{We visually inspect our sample after applying the UnWISE mask, finding a small remaining number of regions where the AGN distribution mirrors the pattern of large-scale diffraction spikes.}{After applying the aforementioned masks, we visually inspect the sky distribution of our sample using {\tt TOPCAT} \citep{2005ASPC..347...29T} to check for remnant artifacts not captured by the masks. We find a small number of remaining regions which suffer from egregious bright star artifacts, where the quasar candidate sky distribution mirrors the pattern of large-scale ($>1$ degree) diffraction spikes.} As our candidate quasars are selected on their very red mid-infrared colors, these regions appear to be affected by the reddest and brightest infrared stars in the sky. We therefore find that using stars selected in the W3 band allows us to effectively mask these regions. We query the AllWISE catalog for sources brighter than W$3 \ [\mathrm{Vega}] < -1.5$ (of which there are only $\sim 200$ sources in the extragalactic sky), and mask regions around them. We find that using a mask diameter in degrees equal to the absolute value of the W3 Vega magnitude effectively removes the affected regions, as brighter stars tend to affect larger areas. We note that this choice represents a convenient rather than optimal solution. Finally, we observe an increase of the density of quasar candidates without optical counterparts \replaced{in}{within a few degrees of} the Ecliptic poles, where the \textit{WISE} imaging is deepest. We thus \added{conservatively} apply a mask within 10$^\circ$ of each Ecliptic pole.

\begin{figure}
    \centering
    \includegraphics[width=0.45\textwidth]{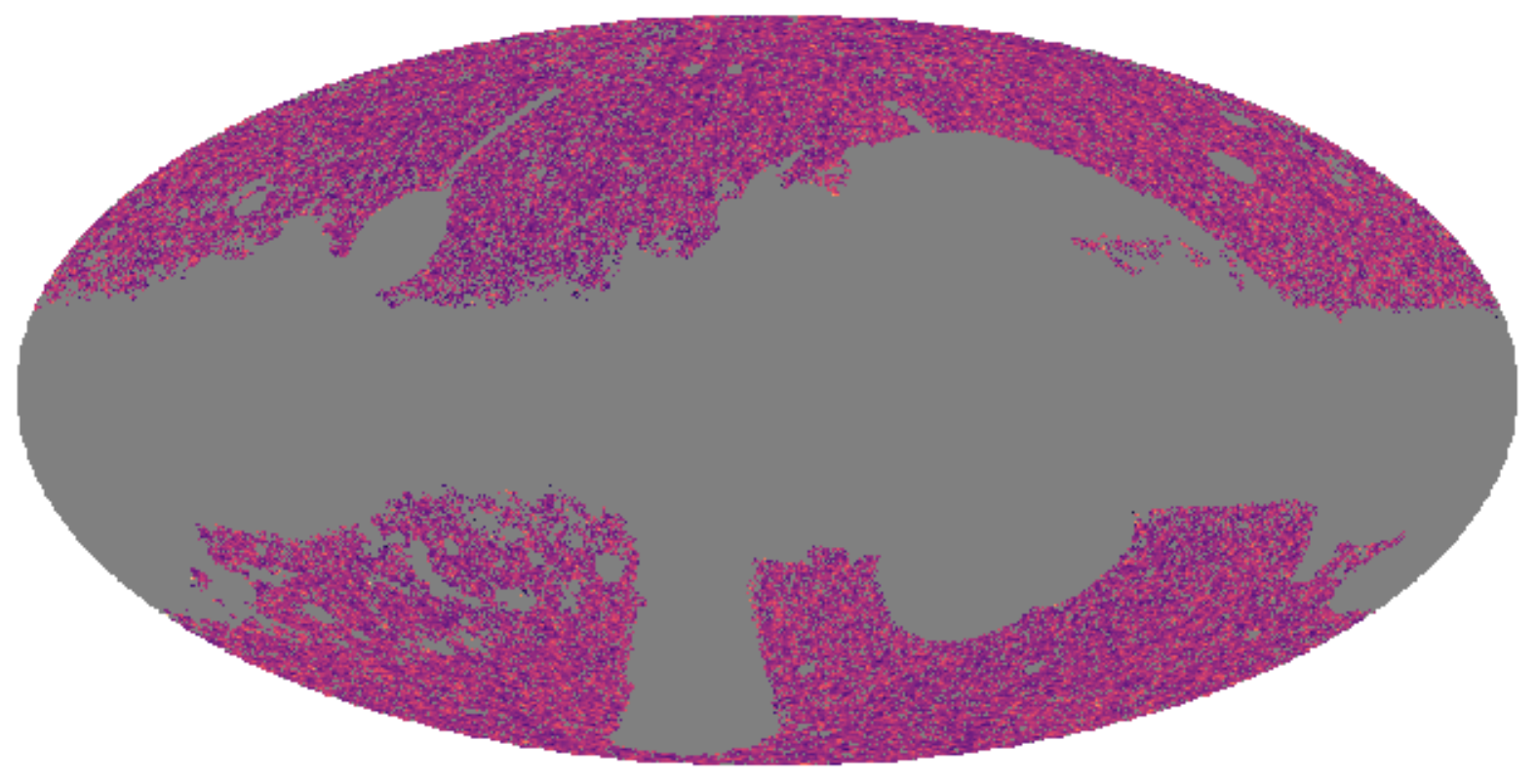}
    \caption{\added{The sky density of $\sim 1.4$ million \textit{WISE} quasar candidates over an effective footprint of $\sim 15,000$ deg$^2$, shown in Galactic coordinates. The density map has been smoothed with a 30$\arcmin$ FWHM Gaussian for visual clarity. The combined mask, shown in gray and summarized in $\S$\ref{sec:masking} shows significant complexity on both large and small scales, resulting in a quasar sample with sufficiently reduced angular systematics to perform clustering measurements.}}
    \label{fig:skydensity}
\end{figure}

\added{We summarize the sample selection and masking as follows. The \citet{2018ApJS..234...23A} R90 selection of Eq.\ \ref{eq:select} curates a sample of $\sim$2.3 million \textit{WISE}-detected sources within the $\sim 20,000$ deg$^2$ DESI-LS footprint. We report the percent reduction for each masking step in reference to this original sample as many sources are masked by multiple masking steps. 5$\%$ of the original sources are excised by the contamination mask  generated following \citet{2018ApJS..234...23A} ($\S$\ref{sec:sample}). 10$\%$ of the original DESI-LS footprint is masked by our optical imaging depth requirement, while 6$\%$ of the area is vetoed by our reddening mask. The UnWISE bitmask flagging procedure removes $6\%$ of sources, and the lost-area mask removes $3\%$. The bright infrared star mask removes $1\%$ of the footprint, while the ecliptic pole mask removes $2\%$. After removing the 12$\%$ of likely contaminants with galaxy-like colors ($\S$\ref{sec:purity}) and the $1\%$ of stellar interlopers with large proper motions ($\S$\ref{sec:sample}), we are left with a sample of $\sim 1.4$ million quasar candidates over an effective area of $\sim$15,000 deg$^2$. We display the final combined mask atop the sky density of quasar candidates as a visual representation of this summary in Figure \ref{fig:skydensity}.}

We apply all above masks to both the data and random catalogs required to perform a clustering measurement. We also mask the \textit{Planck} lensing map with the mask provided alongside the \citet{2020A&A...641A...8P} data release.

\subsection{Weighting Scheme}
\label{sec:weights}

After applying the aforementioned masks to the data and random catalogs, we develop weights to further correct for large-scale angular systematics in the density of quasars \citep[e.g.,][]{2017MNRAS.469.4630D}. These weights will be used in both the lensing and the clustering analyses.

We might expect our sample density to vary with Galactic latitude due to increasing stellar density near the Galactic plane, or Ecliptic coordinates reflecting the \textit{WISE} scanning pattern. Furthermore, by using optical data to bin the sample into obscured and unobscured objects, we might expect each subsample's density to vary with optical imaging depth. We measure the quasar density as a function of each of these variables and find the variation is largest for Ecliptic latitude and logarithmic optical imaging depth. We thus generate two-dimensional histograms of each sample along these dimensions and assign weights as the ratio of data points to random points in each bin. We have subsampled the random catalog using the assigned weights such that the randoms match the data, and use these random catalogs in the correlation function measurements. Alternatively, for the CMB lensing cross-correlation, we assign the inverse of the weights to the data points rather than the randoms, using these weights when constructing the quasar overdensity field.

\subsection{Angular Correlation Functions}
\label{sec:2pcf}

Without redshift information for each of the objects in our quasar sample, we are limited to calculating angular correlation functions, which measure the clustering projected onto the surface of the sky. The two-point angular autocorrelation function $w(\theta)$ is defined as the excess probability --- above that expected from an unclustered Poisson distribution --- of finding a pair of objects at angular separation $\theta$ \citep{1980lssu.book.....P}. We estimate the angular correlation function in this work using the \citet{1993ApJ...412...64L} estimator:

\begin{equation}
    w(\theta) = \frac{DD(\theta) - 2 DR(\theta) + RR(\theta)}{DD(\theta)}, 
\end{equation}
where $DD$, $DR$, and $RR$ are weighted counts, normalized by number density, as a function of separation, for data-data pairs, data-random pairs, and random-random pairs, respectively. 

We measure the angular correlation function using {\tt Corrfunc} \citep{2020MNRAS.491.3022S} in 15 logarithmically-spaced bins between scales of $10^{-2.5} < \theta < 10^{-0.25}$ degrees. This range probes both regimes in which the one-halo term dominates, as well as the two-halo term, while avoiding the $\sim 6 \arcsec$ resolution limit of \textit{WISE} and also the regime at large scales where the Limber approximation begins to break down at the $>10 \%$ level, which is $\gtrsim 1^{\circ}$ for our sample's redshift distribution \citep{2007A&A...473..711S}.

To estimate uncertainties on the correlation functions, we perform a bootstrap resampling of the data \citep[e.g.,][]{1982jbor.book.....E, 2009MNRAS.396...19N}. We divide the sample footprint into 30 equally-sized patches using the $k$-means clustering algorithm within the {\tt scikit-learn} package \citep{2012arXiv1201.0490P}. We then randomly draw from these patches with replacement, and recalculate a correlation function with the data and random points within these patches. This process is repeated 500 times and the variance across realizations is taken to be the variance of our measurement. We find that this error estimation agrees well with analytic Poisson errors in all cases on these scales \citep[e.g.,][]{2006ApJ...638..622M}.

\subsection{Lensing Cross-Correlations}

We also calculate the cross-power spectra $C_{\ell}^{\kappa q}$ of the \textit{WISE} quasar overdensity fields with the CMB lensing convergence field measured by the \textit{Planck} satellite. First, a quasar overdensity map is produced at the same resolution as the lensing map by performing a weighted count of sources in each cell, where the weights correct for the angular systematics discussed in $\S$\ref{sec:weights}. We correct this density field for effects of sub-pixel masking by dividing the density by the fractional-area-lost map. Finally, we convert this into a fractional overdensity map:
\begin{equation}
    \delta_{q} = \frac{\rho - \langle \rho \rangle}{\langle \rho \rangle}
\end{equation}

We estimate the cross spectrum between the two maps using the psuedo-$C_{\ell}$ \citep[e.g., ][]{1973ApJ...185..413P} algorithm {\tt MASTER} \citep[][]{2002ApJ...567....2H}, as implemented in the {\tt NaMaster} package \citep{2019MNRAS.484.4127A}. This algorithm allows for fast and nearly-unbiased estimation of angular power spectra in the presence of sky masks. We measure the quasar-lensing cross-spectrum in 10 logarithmically-spaced bins of angular multipole moment ($\ell$) from $100 < \ell < 1000$. When fitting models to the observed spectrum, we bin the theoretical curves using the same mode-coupling matrix and bandpower binning scheme as used for the data, though in all figures we show the unbinned model spectra for simplicity. 

To estimate uncertainties on the cross-spectra, we utilize simulated noise maps released as part of the Planck lensing data product. After generating 60 noise maps in the same manner as the data, we measure the cross-spectrum between the quasar overdensity field and each of the noise maps, taking the variance as our uncertainty estimate.

\section{Modeling}
\label{sec:model}

\subsection{Modeling Correlation Functions}

We model the correlation functions in a halo occupation distribution (HOD) framework \citep[e.g.,][]{2002ApJ...575..587B}. To first order, the HOD $\langle N (M)\rangle$ is the mean number of quasars belonging to halos of mass $M$, decomposed into contributions from quasars at the centers of halos, $\langle N_c(M) \rangle$, and secondary or `satellite'' quasars belonging to the same halos $\langle N_s(M) \rangle$:

\begin{equation}
    \langle N\rangle = \langle N_c \rangle + \langle N_s \rangle
\end{equation}

The average number density of quasars is then an integral of the HOD over the halo mass function $dn/dM (M,z)$, for which we adopt the model of \citet{2008ApJ...688..709T}:

\begin{equation}
    n_q(z) = \int dM \langle N \rangle \frac{dn}{dM}
\end{equation}

We adopt the HOD model developed by \citet{2007ApJ...667..760Z} and \citet{2011ApJ...736...59Z}, which models the occupation of central objects as a softened step function:

\begin{equation}
    \langle N_c\rangle = \frac{1}{2} \left[1 + \mathrm{erf}\left(\frac{\mathrm{log}_{10}(M / M_{\mathrm{min}})}{\sigma_{\mathrm{log}_{10}M}}\right)\right]
\end{equation}
where \textbf{erf} is the error function, $M_{\mathrm{min}}$ is the characteristic minimum halo mass required to host a quasar, and $\sigma_{\mathrm{log}_{10}M}$ is the softening parameter. The satellite HOD is given by:

\begin{equation}
\label{eqn:satellite}
    \langle N_s \rangle = \Theta(M-M_0) \left(\frac{M - M_0}{M_1}\right)^{\alpha}
\end{equation}
where $\Theta$ is the Heaviside step function, $M_0$ is the minimum mass to host a satellite quasar, and $M_1$ is the mass at which the term transitions to the power-law form. \added{We note that while the  an HOD model of this form has often been used in halo occupation studies of AGNs and quasars \citep[e.g.,][]{2012MNRAS.419.2657C, 2012ApJ...755...30R, 2019MNRAS.487..275G, 2020MNRAS.497..581A}, and thus we adopt this model for the sake of comparison to results from the literature.}

This HOD model has five free parameters, which our present data will not be able to simultaneously constrain. We thus simplify the model as follows. We set $M_{0} = M_{\mathrm{min}}$, such that the minimum halo mass required to host a satellite is the same as that required to host a central quasar. We fix the relatively unimportant parameter governing the softness of the step function to $\sigma_{\mathrm{log}_{10}M}$ = 0.4 dex. Finally, we fix $M_{1} = 12 \times M_{\mathrm{min}}$, motivated by the AGN HOD simulation results of \citet{2019MNRAS.487..275G} relying on the empirical accretion rate distributions of \citet{2017MNRAS.471.1976G} and \citet{2018MNRAS.474.1225A} for luminous AGN. We note that fixing $M_1$ to be the same for obscured and unobscured quasars may not be valid. However, $M_1$ and $\alpha$ are degenerate with one another, roughly driving the relative strength of the one-halo term and thus the fraction of quasars which are satellites. Breaking the degeneracy between $\alpha$ and $M_1$ is not possible with present data. Importantly however, breaking this degeneracy in Eq.\,\ref{eqn:satellite} is not critical to our analysis or interpretation, as we are primarily concerned with deriving the satellite fraction via $\langle N_s \rangle$. Thus, the two free parameters in our model are $M_{\mathrm{min}}$ and $\alpha$. These parameters roughly govern the two-halo term and the one-halo term, respectively.

With HOD parameters specified, the derived halo properties of interest can be expressed. These include the effective bias:

\begin{equation}
    b_{\mathrm{eff}} = \int dz \frac{1}{n_q} \frac{dN}{dz}  \int dM \frac{dn}{dM}
    \langle N\rangle  b_h(M,z) 
    \label{eq:beff}
\end{equation}
and the satellite fraction:
\begin{equation}
    f_{\mathrm{sat}} = \int dz \frac{1}{n_q} \frac{dN}{dz}  \int dM \frac{dn}{dM}  \langle N_s\rangle 
\end{equation}

The effective halo mass $M_{\mathrm{eff}}$ is computed by solving for the mass which would result in the effective bias (Eq.\ \ref{eq:beff}) when integrated over a sample's redshift distribution.

For a given HOD, the power spectrum of the quasar overdensity can be written as the sum of power contributed by pairs of quasars within the same halos (the one-halo term, $P_{1h}(k, z)$) and pairs of quasars between distinct halos (the two-halo term, $P_{2h}(k, z)$).

The one-halo term can be decomposed into pairs between satellites and pairs between satellites and central quasars. The satellites are prescribed to follow the density profile of halos. The function $\tilde{u}(k, M, z)$ is the Fourier transform of the dark matter density profile, for which we adopt the ``NFW'' \citep[][]{1997ApJ...490..493N} model. We use the analytic solution of this profile's transform given by \citet[][]{2001ApJ...546...20S}:

\begin{equation}
    P_{1h} = \frac{1}{n_q^2} \int dM \frac{dn}{dM} [2 \tilde{u} \langle N_c N_s \rangle + \tilde{u}^2 \langle N_s (N_s - 1)\rangle]
\end{equation}
We assume the satellites follow Poisson statistics such that $\langle N_s (N_s - 1)\rangle = \langle N_s \rangle^2$. Setting $M_0 = M_{\mathrm{min}}$ (see above) also imposes the ``central condition'' \citep{2021A&C....3600487M}, which stipulates that a halo can host satellite quasars only if it hosts a central quasar, such that $\langle N_c N_s \rangle = \langle N_s \rangle$. 

The halo density profile transform $\tilde{u}$ approaches unity as $k \rightarrow 0$, such that the one-halo power spectrum approaches a constant at large scales, adding spurious power which can dominate over the two-halo term. We suppress this unphysical power by modifying the one-halo power spectrum with the ad-hoc correction of \citet{2021MNRAS.502.1401M}, with a characteristic damping scale $k^* = 10^{-2} \ h$/Mpc:

\begin{equation}
    P_{1h} \rightarrow P_{1h} \times \frac{(k/k^*)^4}{1 + (k/k^*)^4}
\end{equation}

The two-halo term is given by the matter power spectrum $P_m(k,z)$ multiplied by a bias factor:

\begin{equation}
    P_{2h} = P_m \left[\int dM \frac{dn}{dM} b_h(M, z) \left(\langle N_c\rangle + \tilde{u} \langle N_s\rangle \right) \right]^2
\end{equation}

A known problem in the halo model framework is the underprediction of power compared to $N$-body simulations in the ``quasi-linear regime,'' the transition scales between which the one and two halo terms dominate \citep[e.g., ][]{2014JCAP...08..028F, 2015MNRAS.454.1958M}. This arises due to a natural breakdown of linear perturbation theory as well as halo exclusion effects, though self-consistently modeling these effects is an active area of research. Here, we adopt the empirical function of \citet{2015MNRAS.454.1958M}:

\begin{equation}
    P_{qq} = \left[(P_{1h})^\beta + (P_{2h})^\beta \right]^{1/\beta} ,
\end{equation}
which remedies this effect by smoothing the power spectrum in the transition region. We adopt the best-fit value of $\beta = 0.719$ from \citet{2021MNRAS.502.1401M}.

With a HOD power spectrum specified, we can compute an angular correlation function over the redshift distribution of our sample using the \citet{1953ApJ...117..134L} approximation \citep[e.g., ][]{1980lssu.book.....P, 1991MNRAS.253P...1P, 2017MNRAS.469.4630D}:

\begin{equation}
    w(\theta) = \int dz \frac{H(z)}{c} \left(\frac{dN}{dz}\right)^2  \int \frac{dk \ k}{2 \pi} P_{qq} \  J_{0}[k \theta \chi(z)]
    \label{eq:ang_cf}
\end{equation}
where $J_{0}$ is the zeroth-order Bessel function of the first kind. The $k$-space integral of Eq.\,\ref{eq:ang_cf} is simply a Hankel transform of the power spectrum, for which we utilize the {\tt FFTLog} algorithm \citep{1978JCoPh..29...35T, 2000MNRAS.312..257H}.

Thus, given two free HOD parameters of $M_{\mathrm{min}}$ and $\alpha$ along with a source redshift distribution, a model angular correlation function may be calculated. We fit the observed correlation functions using a Markov chain Monte Carlo (MCMC) method, as implemented in the {\tt emcee} package \citep{2013PASP..125..306F}.

\subsection{Modeling the Lensing Cross-Spectrum}
\label{sec:lens_model}

We also model the cross-spectrum between matter overdensity and CMB lensing convergence. The quasar overdensity and lensing fields are both projections of three-dimensional density fields onto the plane of the sky, and thus the cross-spectrum is given by a line-of-sight integral of the matter power spectrum $P(k, z)$ over the two respective projection kernels under the \citet{1953ApJ...117..134L} approximation with the first order correction \citep[$l \rightarrow l+1/2$; ][]{2008PhRvD..78l3506L}:

\begin{equation}
    C^{\kappa q}_{\ell} = c \int dz \ \frac{W^{\kappa}(z) W^{q}(z)}{\chi^2(z) H(z)}  P\left(k = \frac{\ell + 1/2}{\chi(z)}, \ z\right)
\end{equation}

Here, $c$ is the speed of light, $\chi(z)$ is the comoving distance, and $H(z)$ is the Hubble parameter. We generate the matter power spectra using the analytic form of \citet{1998ApJ...496..605E}. The CMB lensing kernel, a measure of the efficiency of lensing by structure (when multiplied by $d\chi/dz = c/H(z)$) as a function of redshift is given by \citep[e.g.,][]{2000ApJ...534..533C, 2003ApJ...590..664S}:

\begin{equation}
    W^{\kappa}(z) = \frac{3}{2} \Omega_{m, 0} \left(\frac{H_0}{c}\right)^2 (1+z) \chi(z) \frac{\chi_{\mathrm{CMB}} - \chi(z)}{\chi_{\mathrm{CMB}}}
    \label{eq:lenskern}
\end{equation}

The quasar overdensity kernel is in turn given by:

\begin{equation}
    W^{q}(z) = b_h \frac{H(z)}{c} \frac{dN}{dz}
    \label{eq:qso_kern}
\end{equation}
where $dN/dz$ is the normalized redshift distribution of the lenses and $b_h$ is the linear halo bias. 

We have elected to fit the lensing spectra as linearly biased with respect to dark matter as we have found that our present data are unable to constrain the two-parameter HOD model introduced in the previous section. In particular, the Planck lensing map is noisiest at the small scales (large $\ell$-modes) where the one-halo term dominates. Therefore, we instead fit the lensing spectra by assuming an effective halo mass as the single free parameter and inserting the mass-bias relation of \citet{2010ApJ...724..878T} into Eq.\,\ref{eq:qso_kern}.

\section{Results} \label{Results}
\label{sec:results}

\begin{figure}
    \centering
    \includegraphics[width=0.45\textwidth]{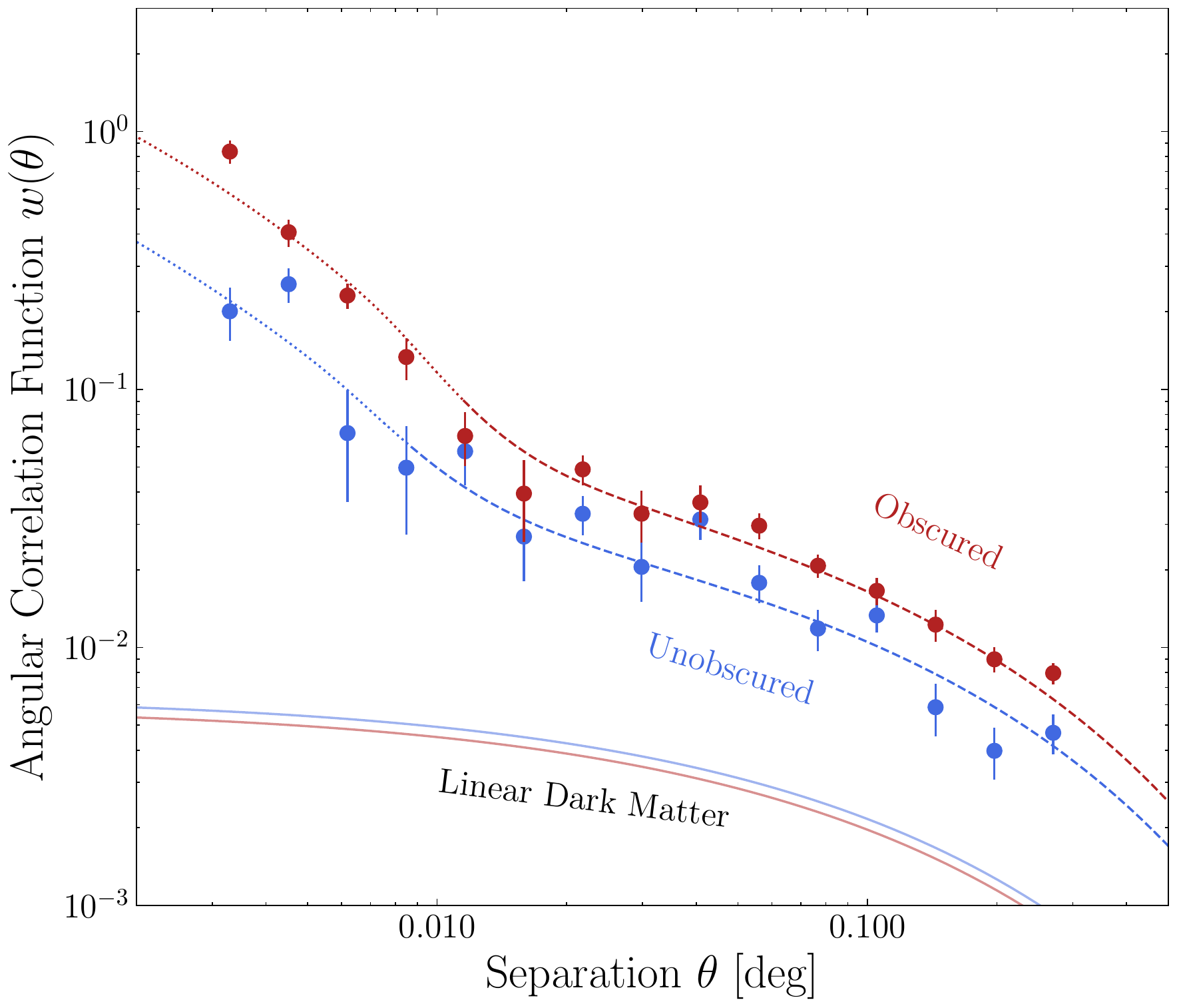}
    \caption{The angular autocorrelation functions of obscured and unobscured quasars are shown with red and blue markers, respectively. The model dark matter correlation functions for the corresponding redshift distributions are shown with solid lines. Finally, the best HOD model fits are shown with dashed/dotted lines. The dotted line indicates where the one-halo term dominates, while the dashed line indicates two-halo term domination. Obscured quasars cluster more strongly than their unobscured counterparts, implying they are a more biased tracer of matter and occupy more massive halos.}
    \label{fig:cfs}
\end{figure}

The results of the angular autocorrelation measurements are displayed in Figure \ref{fig:cfs}. It is apparent that obscured quasars cluster significantly more strongly than their unobscured counterparts (across a similar redshift range), implying that obscured quasars are more biased tracers of matter and occupy more massive dark matter halos. The data for both samples are very well fit by our two parameter HOD model. However, the relative strength of the one-halo term is similar for the two populations, implying similar satellite fractions.

\begin{figure*}
    \centering
    \includegraphics[width=0.9\textwidth]{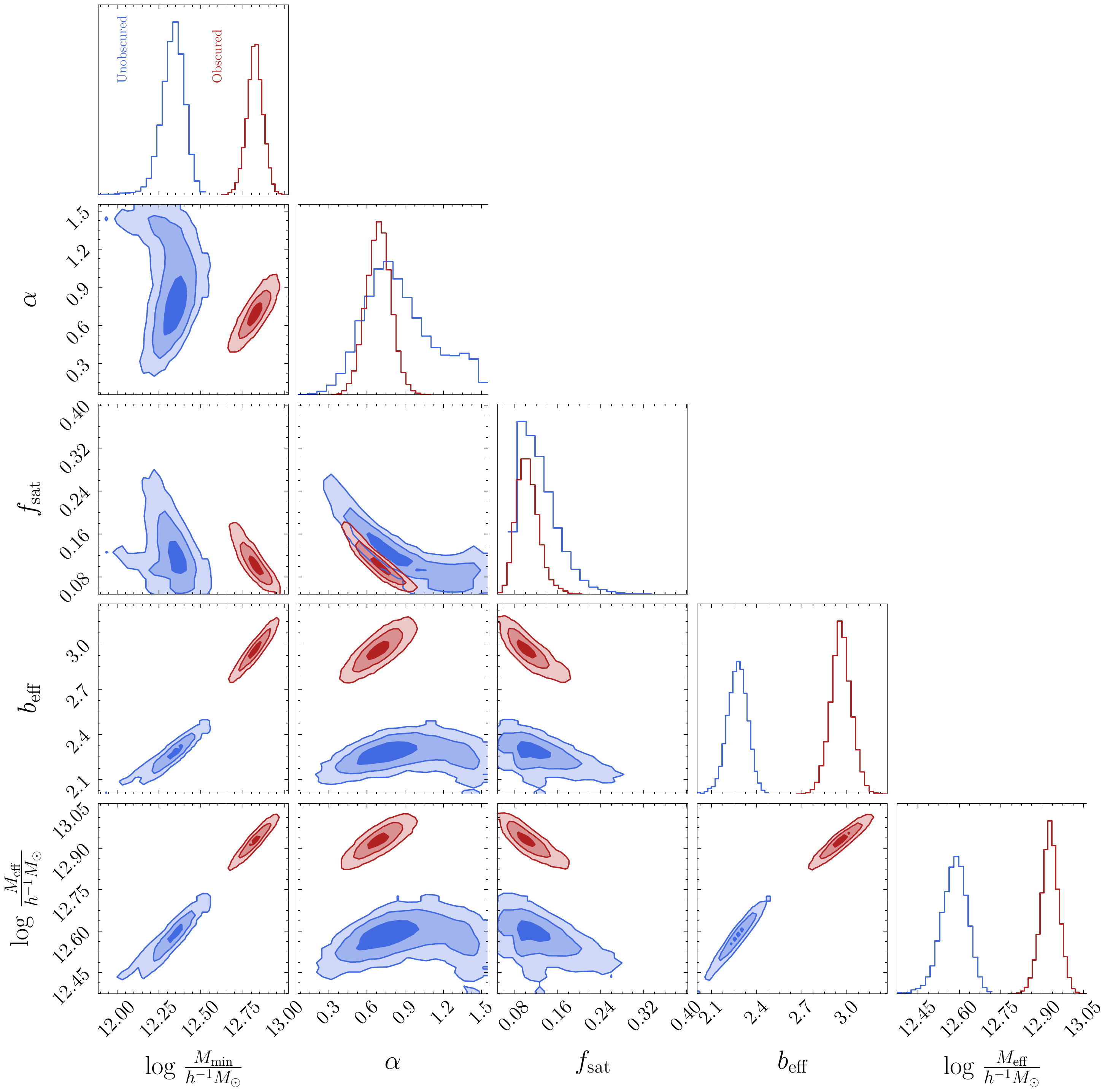}
    \caption{The posterior distributions of our two-parameter HOD model from MCMC fits to the angular autocorrelation functions of obscured and unobscured quasars, shown with red and blue contours, respectively. It is clear that the host halo properties differ significantly, with obscured quasars requiring a larger minimum host halo mass, implying that the effective mass and bias are also larger. However, we do not detect a difference in the one-halo term index $\alpha$, or the satellite fraction $f_{\mathrm{sat}}$.}
    \label{fig:corner}
\end{figure*}

The posterior distributions of the HOD parameters from the MCMC fits to the correlation functions are displayed in Figure \ref{fig:corner}. It is clear that obscured quasars occupy their host halos in a signifcantly different manner than unobscured quasars. The minimum mass required to host an obscured quasar is higher, implying both larger effective biases and host halo masses. In particular, obscured quasars appear to occupy effective halos of $ \sim 10^{12.9} h^{-1} M_{\odot}$, while unobscured quasars occupy halos of $ \sim 10^{12.6} h^{-1} M_{\odot}$. However, we do not detect a difference in the satellite power law index $\alpha$, nor the derived satellite fraction, finding that $\sim 5-20\%$ of obscured and unobscured quasars are satellites within their halos. A non-detection of a difference in the satellite fraction is in contrast to the study of \citet{2018MNRAS.477...45M}. We thus find from angular clustering measurements that obscured quasars occupy more massive halos than unobscured systems, but are not more likely to be satellites within their halos. 

\begin{figure}
    \centering
    \includegraphics[width=0.45\textwidth]{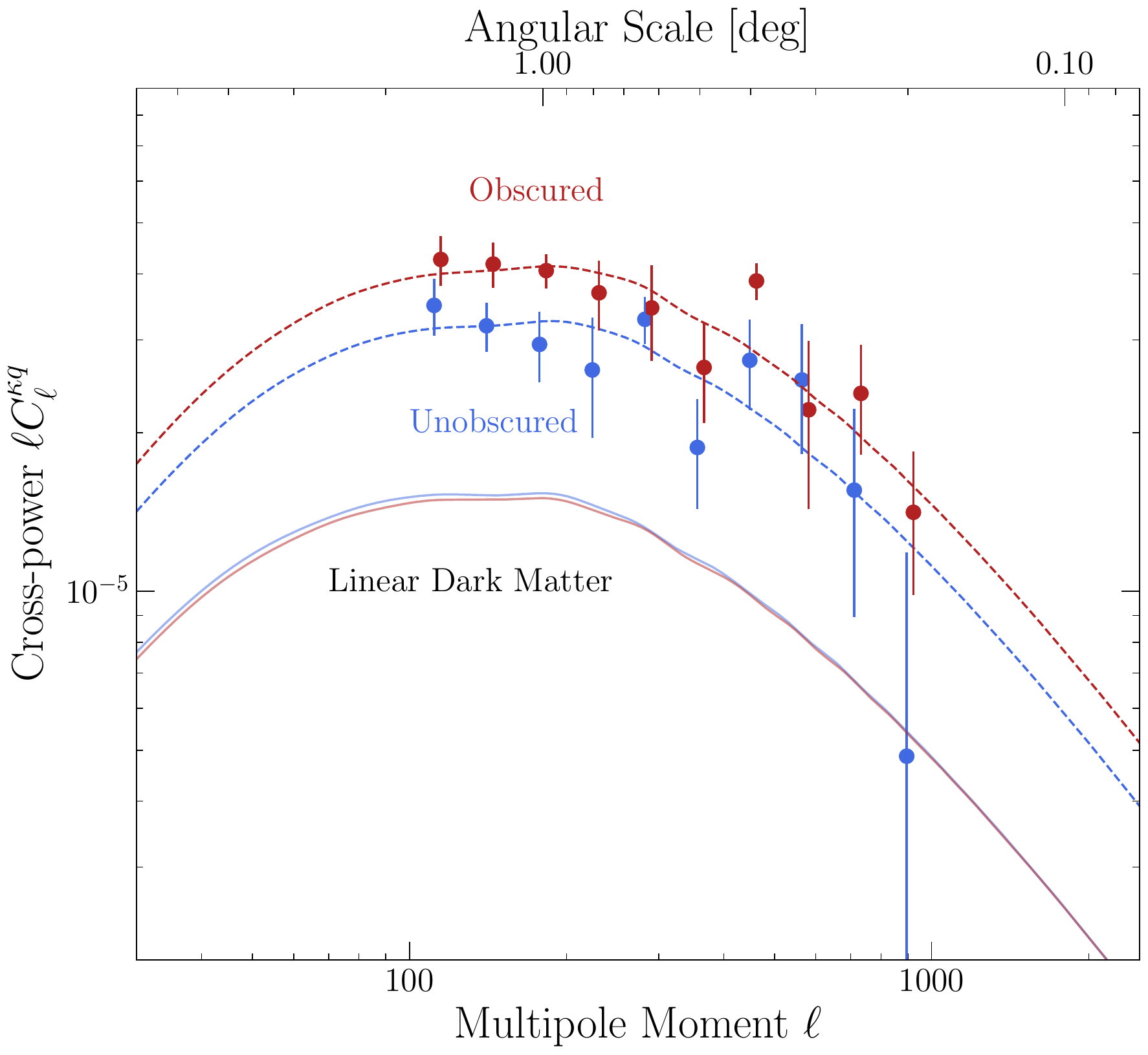}
    \caption{The cross-power spectra (multiplied by $\ell$ to reduce dynamic range) between obscured/unobscured quasar overdensity and Planck CMB lensing convergence $\kappa$ are shown with red and blue markers. The model spectra of linear dark matter for the corresponding redshift distributions are shown using solid lines, while the best model fits are shown with dashed lines. Obscured quasar density correlates more strongly with lensing convergence, implying that obscured quasars are more biased tracers of matter and occupy more massive halos. }
    \label{fig:lensing}
\end{figure}

We also investigate the host halo properties of quasars as a function of obscuration using an independent technique, by calculating cross-correlations of quasar overdensities with the CMB lensing convergence map measured by \textit{Planck}. We display the result of this in Figure \ref{fig:lensing}. In agreement with the clustering analysis, we find that obscured quasars are a significantly more biased tracer of dark matter, again implying more massive host halo environments. We are unable to constrain the one-halo term, and thus satellite fraction with present data, as the \textit{Planck} CMB lensing map is noisiest at small scales, as discussed in $\S$\ref{sec:lens_model}. Therefore, we fit the observed spectra with linearly-biased dark matter models, which provide good fits to the data.

\begin{figure}
    \centering
    \includegraphics[width=0.45\textwidth]{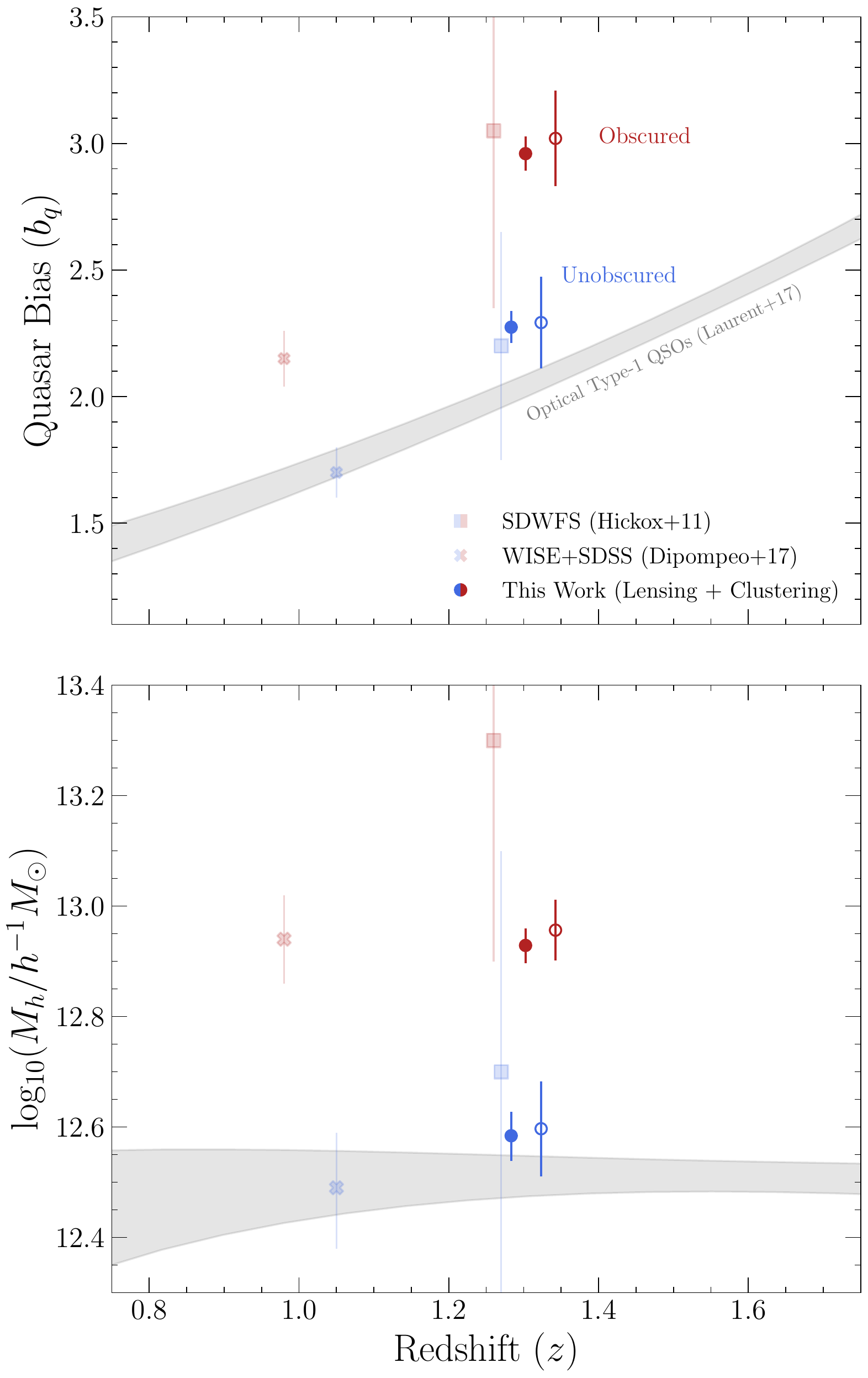}
    \caption{The bias (top panel) and effective halo mass (bottom panel) of obscured and unobscured quasars. We show the results for obscured quasars in red, and unobscured quasars in blue. Our angular clustering results are shown with closed circles, while the CMB lensing results are displayed with open circles, offset horizontally for visual clarity. We also show clustering measurement results of infrared quasars at similar effective redshifts from \citep{2011ApJ...731..117H} and \citep{2017MNRAS.469.4630D}, as well as for SDSS Type-1 quasars \citep{2017JCAP...07..017L}. }
    \label{fig:bias}
\end{figure}

We summarize our results in Figure \ref{fig:bias} and tabulate them in Table \ref{table:result}. This shows most strikingly that the implied effective halo masses from clustering and CMB lensing analyses are in excellent agreement, with both showing that obscured quasars occupy significantly more massive effective halos. We show that infrared-selected unobscured quasars occupy similar effective halos as optically-selected spectroscopic Type-1 quasars from eBOSS \citep[e.g.,][]{2017JCAP...07..017L}. We also show that our results are consistent with those of \citet{2017MNRAS.469.4630D}, demonstrating that the enhanced clustering of \textit{WISE}-selected obscured quasars persists both with improved precision and at higher redshifts.

\begin{deluxetable}{cccc}
\label{table:result}
\tablehead{\colhead{Analysis} & \colhead{Sample} & \colhead{$b_q$} & \colhead{log$_{10}(M_h / h^{-1} M_{\odot}$}}
\startdata
\hline
Clustering \\
\hline
 & Unobscured & $2.27 \pm 0.06$ & $12.58 \pm_{0.05}^{0.04}$ \\
 & Obscured & $2.96 \pm 0.07$ & $12.93 \pm 0.03$ \\
\hline
Lensing \\
\hline
 & Unobscured & $2.29 \pm 0.18$ & $12.60 \pm 0.09$ \\
 & Obscured & $3.02 \pm 0.19$ & $12.96 \pm 0.06$
\enddata
\caption{A tabulation of the measured halo biases and the corresponding effective halo masses for obscured and unobscured quasars. We display results from angular clustering analyses as well as from CMB lensing, which are in excellent agreement and show obscured quasars occupy more massive halos. The contents of this table are visually represented in Figure \ref{fig:bias}.}
\end{deluxetable}

\section{Discussion}

\label{sec:discussion}

In $\S$\ref{sec:results}, we have shown that \textit{WISE}-selected obscured quasars occupy significantly more massive halos than unobscured quasars do. In this section, we will discuss possible systematics in these measurements and speculate on interpretations of this observed halo occupation difference.

\subsection{Redshift Systematics}

Inferring the host halo properties of a sample from angular statistics is inherently sensitive to systematics in the sample's redshift distribution, with the uncertainty in the latter oftentimes dominating over the statistical uncertainty of the clustering measurement \citep[e.g., ][]{2013pss6.book..387C}. As approximately half of our estimation of the obscured quasar redshift distribution relies on photometric redshifts (Figure \ref{fig:zdist}), it is important to quantify whether our result could be explained by photometric redshift systematics, as was suggested by the authors of \citet{2016ApJ...821...55M} to explain the similar results of \citet{2014ApJ...789...44D} and \citet{2014MNRAS.442.3443D}. 

\begin{figure}
    \centering
    \includegraphics[width=0.47\textwidth]{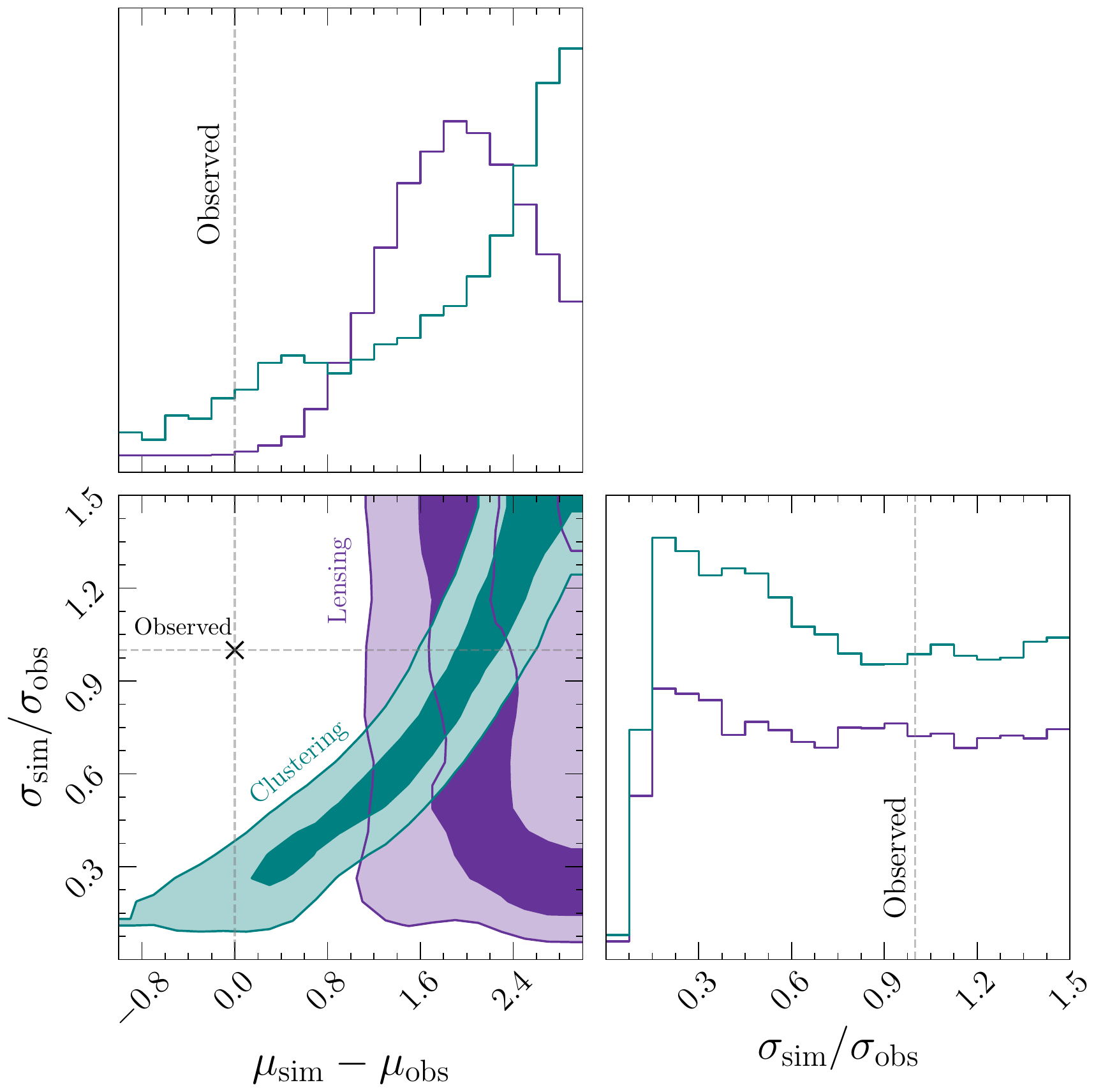}
    \caption{A corner plot demonstrating a test for systematics in our effective halo mass results due to photometric redshift distribution uncertainties. We have approximated the measured photometric redshift distribution of obscured quasars as a Gaussian, then shifted the mean $\mu$ and dispersion $\sigma$ from their observed values until the implied effective halo mass for obscured quasars matches that of unobscured quasars. We show the 68$\%$ and 95$\%$ confidence contours for the redshift distribution properties required to resolve the halo mass discrepancy for the angular clustering and CMB lensing measurements in teal and purple, respectively. In order to resolve the halo mass tension while reproducing both the clustering and lensing signals, the half of our obscured quasar sample in deep fields without spectroscopic redshifts would need to be distributed around $z\approx 3.5$ instead of the value derived from photometric redshifts of $z=1.4$, which we regard as implausible. Our result is thus robust against redshift systematics.}
    \label{fig:systematic_test}
\end{figure}

We test this possibility by exploring the potential obscured quasar redshift distributions which would resolve the observed effective halo mass difference between obscured and unobscured quasars while reproducing our measurements. This was achieved by replacing the observed photometric redshift distribution of obscured quasars with many simulated distributions and refitting the data. First, the obscured quasar photometric redshift distribution of Figure \ref{fig:zdist} was fit as a normal distribution, finding a mean and dispersion of $\mu_{\mathrm{obs}} = 1.4$, $\sigma_{\mathrm{obs}} = 0.5$. We then vary these parameters, shifting the mean by $-1 < \mu_{\mathrm{sim}} - \mu_{\mathrm{obs}} < 3$ and varying the dispersion between $0.1 < \sigma_{\mathrm{sim}} / \sigma_{\mathrm{obs}} < 1.5$. Redshifts are randomly drawn from this new distribution, and only values $z>0$ are retained. Distributions with $>10\%$ of the cumulative distribution function below zero are rejected outright. These simulated redshifts are finally recombined with the observed spectroscopic redshifts. We then refit the obscured quasar clustering and lensing signals (on scales dominated by the two-halo term) and determine which combinations of the redshift distribution parameters imply an effective halo mass consistent with that measured for unobscured quasars. The combinations which resolve the tension are explored in an MCMC analysis, for which we display the results in Figure \ref{fig:systematic_test}.

It is clear that the dependence of clustering and CMB lensing measurements on redshift systematics are independent, with clustering being primarily sensitive to the distribution width and the lensing sensitive to the mean. We observe that the only possible configuration which would resolve the halo mass tension while reproducing both the clustering and lensing measurements would be if the obscured quasars without spectroscopic redshifts were distributed around $z\approx 3.5$, which we regard as infeasible. The photometric redshifts stem mostly from \citet{2021A&A...648A...4D}, which reports a typical photometric redshift scatter of $7\%$ and an outlier fraction of $\sim 20\%$ for AGNs. In order to resolve the observed clustering differences, the true outlier fraction for obscured quasars would need to be nearly $100\%$ if these systems were truly narrowly distributed at $z\approx 3.5$. Therefore, we argue that our measurements of halo mass differences are robust against systematics from uncertainties in the redshift distribution. This test highlights the power of combining clustering and lensing measurements to mitigate redshift systematics.

\subsection{Interpretation of Clustering Difference}
A significant observed difference in the host halo properties of matched samples of obscured and unobscured quasars appears to negate the simplest unified model which attributes obscuration solely to viewing angle. An interesting alternative in the context of galaxy evolution is an evolutionary model which posits that obscured and unobscured quasars represent different phases in a coevolutionary scheme between galaxies and their nuclear black holes. \citet{2017MNRAS.464.3526D} proposed a simple evolutionary model in which black hole growth lags behind dark matter halo growth \citep[e.g.,][]{2008AJ....135.1968A, 2008ApJ...681..925W, 2013ARA&A..51..511K}. If obscured quasars represent an early evolutionary phase of rapid SMBH growth and thus their black holes are systematically undermassive with respect to their halos (they have not yet ``caught up'' to the SMBH halo relation), then the obscured subset of a luminosity-limited quasar sample (which is in turn black hole mass-limited for a fixed Eddington ratio distribution) will systematically occupy more massive halos. 

However, it is also possible that such a difference could arise from non-evolutionary effects. The modeling work of \citet{2020ApJ...888...71W} explored whether the observed clustering difference between obscured and unobscured quasars reported by \citet{2017MNRAS.469.4630D} could be explained by such non-evolutionary scaling relations. In particular, they studied whether the results could be understood within the  ``radiation-regulated unification'' schema of \citet{2017Natur.549..488R}, where an observed relationship between AGN Eddington ratio and torus covering factor is interpreted to imply that the radiation pressure from black holes accreting closer to their limits can expel toroidal dust and expose more unobscured sight lines to a potential observer. Therefore, one would expect a reduced incidence of obscured quasars represented at high Eddington ratios, which has been confirmed at low redshift by \citet{2022ApJS..261....9A}. This implies that at fixed luminosity, obscured quasar activity will preferentially be generated by more massive black holes accreting at lower Eddington ratios, which will thus lie in more massive halos according to black hole-halo mass relations \citep[e.g.,][]{2013ARA&A..51..511K}. Such a model could potentially explain the observed host halo difference without necessitating an evolutionary scheme.

\citet{2020ApJ...888...71W} also studied whether the observed clustering difference could be explained by scaling relations between host galaxy stellar mass and ISM column density \citep{2009ApJ...698L.116P, 2017MNRAS.464.4545B, 2017ApJ...850..208W}. As more massive galaxies are observed to contain larger columns of obscuring gas, obscured quasars may be preferentially selected in massive galaxies and thus halos without regard to an evolutionary process.

The modeling work of \citet{2020ApJ...888...71W} showed that the observed halo mass difference of \citet{2017MNRAS.469.4630D} cannot be explained either through radiation-regulated unification nor through scaling relations with galaxy mass while simultaneously reproducing the observed fraction of obscured quasars. With the updated effective host halo masses presented in this work, which are consistent with but roughly twice as precise as the results of \citet{2017MNRAS.469.4630D}, these arguments become even more stringent. We thus favor an evolutionary explanation of the host halo mass difference observed in this work.

However, accurate comparison of the clustering properties between populations of quasars requires controlling across properties such as redshift, quasar luminosity, and host galaxy properties such as stellar mass and star formation rate (SFR). Indeed, several works suggest that AGN clustering properties depend only on the host galaxy property distributions revealed by different AGN selection techniques \citep[e.g.,][]{2016ApJ...821...55M, 2018MNRAS.480.1022Y, 2018ApJ...858..110P, 2020ApJ...891...41P, 2020MNRAS.494.1693K, 2021MNRAS.502.5962A}. In this work, we have shown that \textit{WISE} obscured and unobscured quasars exhibit similar redshift and bolometric luminosity distributions. However, a potential concern is that obscured quasars may appear to occupy more massive halos because of a selection effect in which they are detected in host galaxies differing from the hosts of unobscured systems. Future work on modeling the full spectral energy distributions of \textit{WISE} quasars is required to test whether the observed clustering difference may be understood in terms of host galaxy properties. However, \citet{andonie22} recently performed panchromatic SED modeling of infrared-selected quasars in the COSMOS field, finding no difference in the stellar mass distribution for obscured and unobscured quasars.

\subsection{Non-torus-obscured Population}
\label{sec:nto}
Although a robust trend between quasar obscuration and halo properties rules out the possibility that obscuration is \textit{always} driven by viewing angle with the torus, it does not exorcise the role of nuclear obscuration. Instead, we expect a fraction of obscured quasars to appear obscured due to orientation effects. As the clustering of torus-obscured objects should match that of unobscured systems, the measured clustering of the obscured population in fact represents a lower limit for the population of systems obscured by evolutionary processes, as noted by \citet{2016MNRAS.460..175D}. The true clustering of this ``non-torus-obscured'' (NTO) population then depends on the NTO fraction ($f_{\mathrm{NTO}}$), the proportion of obscured quasars which are obscured by evolutionary effects as opposed to orientation.

We thus investigate the halo properties required of NTO quasars in order to produce the observed clustering as a function of NTO fraction. We adopt the method of \citet{2016MNRAS.460..175D} to compute the implied host halo properties for NTO quasars using our updated measurements. Assuming similar redshift distributions, the bias of the NTO population is given by:
\begin{equation}
    b_{\mathrm{NTO}} = \frac{b_{\mathrm{ob}} - b_{\mathrm{unob}}}{f_{\mathrm{NTO}}} + b_{\mathrm{unob}}
\end{equation} 

\added{Therefore, the smaller the fraction of obscured quasars belonging to the NTO class, the more biased and rare halos NTO quasars must occupy to drive the observed clustering enhancement of all obscured quasars. For a physical interpretation, we use this NTO bias to estimate the minimum host halo mass, halo occupation fraction, and characteristic lifetime of NTO quasars as a function of NTO fraction.} We convert between this NTO bias and a \replaced{effective}{minimum} halo mass required to host a quasar as a function of the NTO fraction $f_{\mathrm{NTO}}$ using: 

\begin{equation}
    b(M > M_{\mathrm{min}}) =  \frac{ \int_{M_{\mathrm{min}}}^{\infty} dM \frac{dn}{dM} b(M)}{ \int_{M_{\mathrm{min}}}^{\infty} dM \frac{dn}{dM}}.
\end{equation}

\begin{figure}
    \centering
    \includegraphics[width=0.45\textwidth]{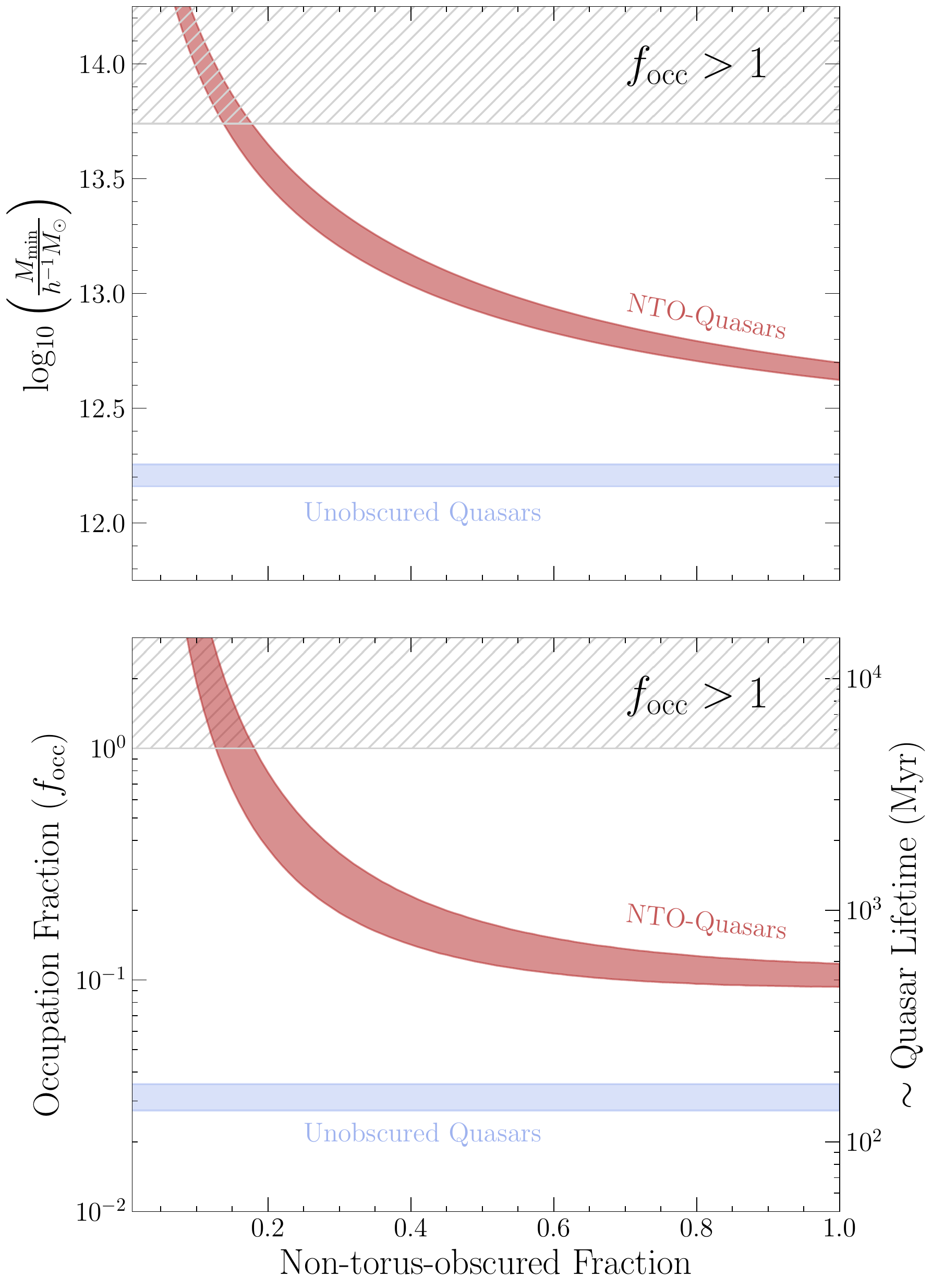}
    \caption{\added{Top panel:} the implied minimum host halo mass for ``non-torus-obscured'' (NTO) and unobscured quasars as a function of the NTO fraction of our sample. \added{Bottom panel: the implied occupation fraction of NTO and unobscured quasars, or the fraction of halos more massive than the corresponding minimum mass which host a quasar of a given type. The right axis shows a characteristic lifetime of quasars in a given phase corresponding to the occupation fraction.} If only a small fraction of obscured quasars belong to the NTO population, the bias and host halo masses of these systems must be very large in order to drive the observed clustering of the entire obscured population. Conversely, this constrains the NTO fraction to be $\gtrsim 15 \%$ in order to avoid unphysically massive host halos, where the quasar occupation fraction exceeds unity (see $\S$\ref{sec:nto}).}
    \label{fig:ntobias}
\end{figure}

\added{We note that this minimum mass differs slightly from that defined in our HOD modeling which includes a softening of the step function. Given a minimum halo mass, we are able to compute an occupation fraction ($f_{\mathrm{occ}}$, the number fraction of halos more massive than a threshold mass which host a given type of quasar) by comparing the space density of quasars with the density of halos more massive than the minimum mass. To estimate the space density of our quasar samples, we integrate the \citet{2020MNRAS.495.3252S} quasar luminosity function (QLF) over the observed $6\,\mu$m luminosity distributions ($\S$\ref{sec:lums}), also integrating over the sample's redshift distribution. Next, we divide the resulting density by two in order to isolate the density of the obscured population (which comprises roughly half of the total, $\S$\ref{sec:data}). For NTO quasars we finally multiply the density by the NTO fraction. Next, we compute the space density of halos as a function of minimum mass by performing mass and redshift integrals over the halo mass function of \citet{2008ApJ...688..709T}.

Finally, we provide estimates of the quasar lifetime, the characteristic time a SMBH is expected to be observable as a quasar of a given phase. Assuming that every halo hosts one SMBH, the quasar lifetime is simply proportional to the halo occupation fraction \citep{2001ApJ...547...27H, 2001ApJ...547...12M}. A majority of our sample quasars lie at $z \approx 0.5-2$, corresponding to a span of $\sim 5$ Gyr of cosmic time. An approximate estimate of the quasar lifetime is thus given by this factor multiplied by the occupation fraction.

We show the results of these calculations in Figure \ref{fig:ntobias}.} It is apparent that if NTO quasars make up a small proportion of the obscured population, their host halos must be very massive in order to drive the clustering of the full obscured population to the observed value. Conversely, we can use this relation to put constraints on the minimum NTO fraction required by our measurements in order to avoid unphysically large host halo masses for these objects. \added{At sufficiently small NTO fractions ($\lesssim 15\%$), the implied host halos of NTO quasars become massive and thus rare enough that the corresponding occupation fraction exceeds unity. This forbidden region is shown in Figure \ref{fig:ntobias} with a gray hatched region.} This appears to be a compelling line of evidence that a non-negligible $(\gtrsim 15\%)$ fraction of obscured quasars are obscured by a mechanism aside from orientation, such as galactic or circumnuclear dust. \added{We also note that our quasar lifetime analysis suggests that if NTO quasars represent a distinct evolutionary phase in the evolution of SMBH growth, the duration of the obscured phase appears greater than the unobscured phase by a factor $\gtrsim 3$}.

\subsection{Possible Connection to Dust-obscured Galaxies}
\begin{figure}
    \centering
    \includegraphics[width=0.45\textwidth]{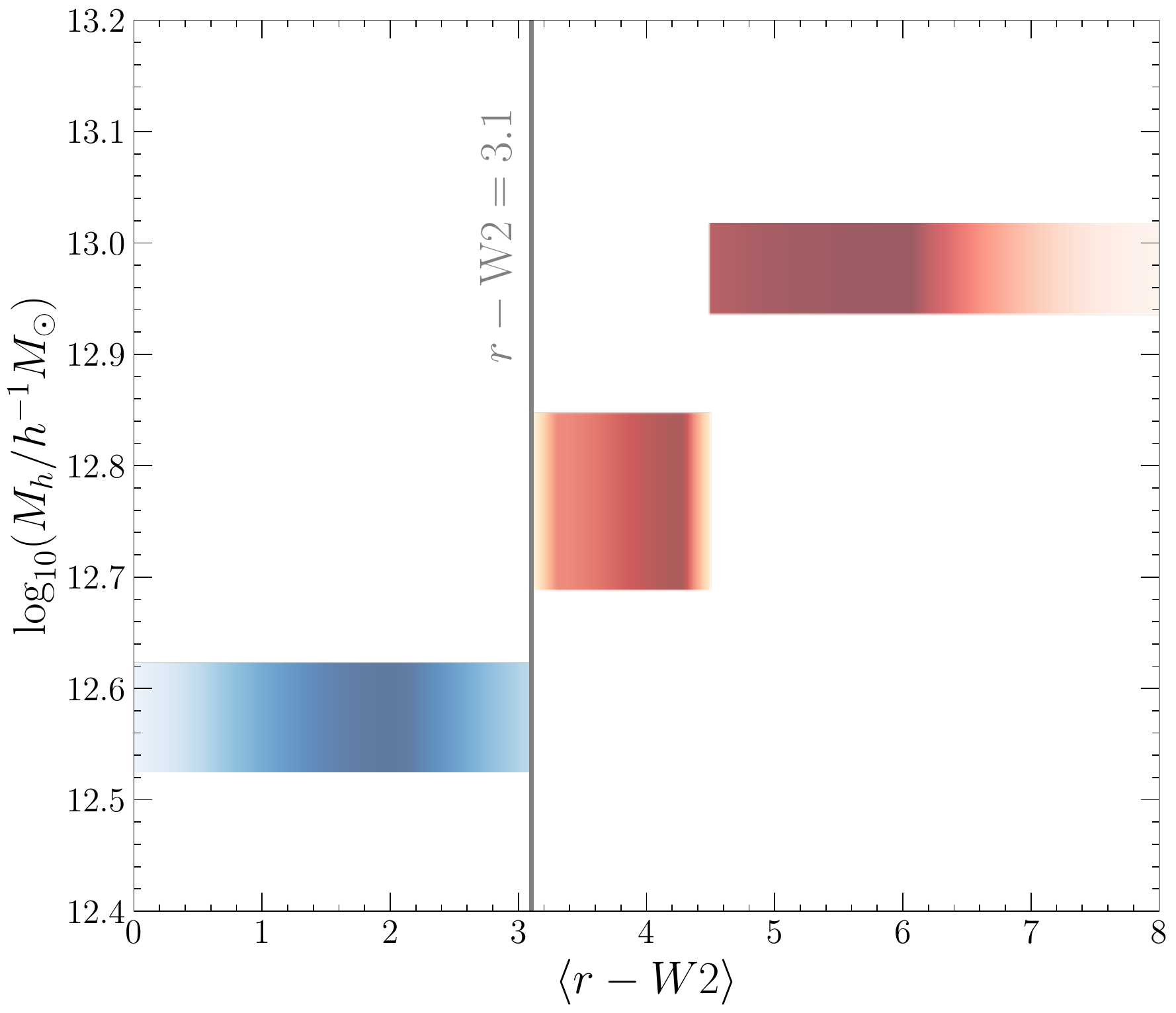}
    \caption{A further exploration of the effective host halo masses of quasars as a function of obscuration. We have now split the obscured sample into two subsets using a color cut ($r$-W2 [AB] = 4.5). We show the effective halo masses of the samples split by optical-infrared color using gradients, where the vertical span represents the 68$\%$ confidence interval, and the opacity of the gradient represents the relative number of sources of a given color. The effective halo mass appears to continue increasing for obscured quasars towards redder colors, suggesting the enhanced clustering may be dominantly driven by the reddest and most obscured subset. }
    \label{fig:dog}
\end{figure}

By matching our quasar catalog to SDWFS 24\,$\mu$m photometry, we find that $\sim 30\%$ of the objects in the the obscured quasar sample would be classified as dust-obscured galaxies \citep[DOGs,][]{2008ApJ...677..943D} with $F_{24} / F_{r} > 1000$. DOGs are ultra-luminous infrared galaxies (ULIRGs) often accompanied by quasar activity at $z = 1-2$, where the optical emission from both stars and SMBH accretion appears heavily attenuated, implying strong galactic-scale absorption. These objects thus are expected to be good candidates for galaxies caught in the violent post-merger evolutionary stage expected to produce obscured quasar activity and star formation in the models of \citet{1988ApJ...325...74S} and \cite{2008ApJS..175..356H}. The DOGs in our sample are expected to be quasar-dominated given that their infrared colors satisfy the \citet{2012ApJ...748..142D} criterion of power-law infrared spectra. We observe that the DOG fraction increases in our obscured quasar sample towards systems with redder $r-$W2 colors. To test whether the enhanced clustering of obscured quasars observed in this work could be driven by the reddest tail of sources with significant overlap with the DOG population, we split the obscured sample into two further subsets and again perform each clustering and lensing analysis as previously presented. We split the obscured population in two using a cut of $r-$W2 = 4.5, which corresponds to the peak of the obscured quasar distribution (Figure \ref{fig:colors}). The subset redder than this cut consists of $\sim 50\%$ DOGs. We do not split the unobscured sample, as \citet{2022ApJ...927...16P} showed that unobscured spectroscopic quasars' clustering is not connected with their $r-$W2 colors. We display the effective halo masses from these analyses as a function of $r-$W2 color in Figure \ref{fig:dog}.

Interestingly, the effective halo mass of \textit{WISE} quasars appears to continue increasing towards more extreme red optical-infrared colors. We find that the results from clustering and lensing again agree, which indicates this trend is robust. The reddest sample appears to occupy halos of effective mass $ \sim 10^{13.0} h^{-1} M_{\odot}$. Meanwhile, the study of \citet{2017ApJ...835...36T} measured the clustering of \textit{WISE}-selected bright DOGs (which tend to be quasar-dominated), finding that they occupy very massive effective halos of $ \sim 10^{13.6} h^{-1} M_{\odot}$. Given a majority of the reddest sample would be selected as DOGs, we speculate that the observed enhanced clustering of obscured infrared-selected quasars may be driven by the reddest and most heavily obscured subset, which are often DOGs and expected to be obscured by host-galaxy material. \added{However, we note that we perform a full HOD fit of each subsample, but do not detect an excess of small scale clustering in the reddest bin as might be expected for a sample dominated by merger systems.}

\subsection{Comparison to Literature Results}

It is important to understand the origin of the wide array of seemingly conflicting results on the clustering of obscured and unobscured AGNs in the literature. Assessing the literature, we broadly find that studies detecting enhanced clustering of obscured quasars tend to be selected in the infrared \citep[][this work]{2011ApJ...731..117H, 2014ApJ...789...44D, 2014MNRAS.442.3443D, 2015MNRAS.446.3492D, 2016MNRAS.456..924D, 2017MNRAS.469.4630D} or ultra-hard X-ray \citep{2018ApJ...858..110P}, while studies which do not detect a difference tend to use softer X-ray selection (e.g., with \textit{Chandra} or \textit{XMM-Newton}) over relatively smaller fields \citep{2009ApJ...701.1484C, 2009A&A...494...33G, 2009A&A...500..749E, 2012MNRAS.420..514M, 2018MNRAS.481.3063K}, though there are exceptions \citep{2013ApJ...776L..41G, 2016ApJ...821...55M, 2018MNRAS.474.1773K}. 

We posit that the diversity of results may be understood by one of two selection effects. In one scenario, quasar selection at certain wavelengths may reveal obscured and unobscured populations which are not matched in host galaxy properties that correlate with dark matter halo mass. In this scenario, it is possible, for instance, that infrared quasar selection may uncover obscured quasars which occupy more massive galaxies than unobscured quasars, which would produce a difference in observed halo mass given that galaxy stellar mass correlates with halo mass. However, the modeling work of \citet{2020ApJ...888...71W} appears to point against this scenario.

Alternatively, it is possible that the effective halo mass of a sample of quasars is driven by an extreme subset of that sample, which are not equally recoverable in all wavebands. In this context, we suggest that the enhanced clustering of obscured quasars may be driven by the most heavily obscured subset, which are able to be selected at infrared and hard X-ray wavelengths but not with soft X-rays. We have shown in Figure \ref{fig:dog} that the clustering of obscured quasars continues to increase towards redder optical-infrared colors and greater overlap with the DOG population, possibly supporting this view. Quasar-dominated DOGs have been shown to often be obscured by Compton-thin or Compton-thick columns \citep[e.g.,][]{2008ApJ...672...94F, 2009A&A...498...67L, 2014ApJ...794..102S, 2015A&A...574L...9P, 2016MNRAS.456.2105D, 2016ApJ...819..111A, 2018MNRAS.474.4528V}, which would prevent their detection in typical soft X-ray surveys. We thus speculate that the seemingly conflicting results in the literature regarding the host halos of obscured and unobscured quasars may be explained if the enhanced clustering of infrared-selected obscured quasars is driven by deeply-buried systems missed by soft X-ray surveys.

\section{Conclusions}

In this paper, we have selected $\sim 1.4$ million quasars over $\sim 15,000 \ \mathrm{deg}^2$ of sky in the mid-infrared with \textit{WISE} and split them into obscured and unobscured samples using optical-infrared colors. We have then probed their host halo properties by measuring their angular clustering statistics as well as the typical gravitational deflection of CMB photons passing through their surrounding large-scale structures. We have interpreted the clustering signals within an HOD framework, finding that the minimum and effective halo masses of obscured quasars are significantly higher than their unobscured counterparts, with obscured quasars occupying effective halos of $ \sim 10^{12.9} h^{-1} M_{\odot}$, while unobscured quasars occupy effective halos of $ \sim 10^{12.6} h^{-1} M_{\odot}$. We however do not detect a difference in the one-halo term, finding that $\sim 5-20\%$ of both obscured and unobscured quasars are satellites within their halos. Interpreting the CMB lensing signals with a linearly biased halo model, we find excellent agreement with the clustering results. We showed that this agreement confirms the halo mass difference is robust against uncertainty in the obscured sample's redshift distribution. We discuss interpretations of the observed clustering difference, and favor an evolutionary explanation for the obscuration of at least some quasars. Finally, we detect a hint that the enhanced clustering is driven by the systems with the most extremely red optical-infrared colors, which have a significant overlap (>$50\%$) with the Dust-obscured Galaxy (DOG) population.

With this work, we have now estimated the halo properties of infrared quasars across the majority of the extragalactic sky. We are therefore likely approaching the limit of the statistical power available to probe the halos of purely \textit{WISE}-selected quasars through angular autocorrelations. Consequently, further investigation of the halo properties of these quasars must utilize other techniques. While the \textit{Planck} map of CMB lensing will remain the only all-sky dataset for the near future, ground-based experiments such as the Atacama Cosmology Telescope, South Pole Telescope, and soon the Simons Observatory and CMB-S4 will improve the lensing precision on smaller scales \citep[e.g.,][]{2017ApJ...849..124O, 2021MNRAS.500.2250D}, and may allow direct measurements of the quasar one-halo term. The weak lensing of background galaxies can also be used to study AGN halo properties \citep[e.g., ][]{2015MNRAS.446.1874L, 2022arXiv220403817L}. Upcoming missions including \textit{Euclid} and the Rubin Observatory will measure the weak gravitational lensing of galaxies at $z > 2$ over a wide area, probing the large-scale structure around the peak of quasar activity at $z = 1-2$. These missions will also detect and measure photometric redshifts for galaxies at $z = 1-2$, enabling cross-clustering measurements to increase the signal-to-noise ratio for the inherently rare population of bright quasars. Finally, spectroscopic surveys of quasars would allow spatial clustering measurements and more precise obscuration criteria, but extending spectroscopy to the most heavily obscured systems will prove challenging due to their inherent optical faintness.

In order to interpret the clustering difference between the obscured and unobscured quasars observed in this work, a comparison of the host galaxy properties through full panchromatic SED modeling \citep[e.g.,][]{andonie22} is required in future work. Such a characterization, combined with the results of this work, will provide a probe of the full halo-galaxy-SMBH connection, testing models of AGN structure and of galaxy-SMBH coevolution.

\acknowledgements

We thank the anonymous reviewer for comments which significantly improved this work. GCP acknowledges support from the Dartmouth Fellowship. DMA thanks the Science Technology Facilities Council (STFC) for support from the Durham consolidated grant (ST/T000244/1). CA acknowledges support from EU H2020-MSCA-ITN-2019 Project 860744 “BiD4BESt: Big Data applications for black hole Evolution STudies. This research has made use of the NASA/IPAC Infrared Science Archive, which is funded by the National Aeronautics and Space Administration and operated by the California Institute of Technology. 

This research uses services or data provided by the Astro Data Lab at NSF's National Optical-Infrared Astronomy Research Laboratory. NOIRLab is operated by the Association of Universities for Research in Astronomy (AURA), Inc. under a cooperative agreement with the National Science Foundation.

The Legacy Surveys consist of three individual and complementary projects: the Dark Energy Camera Legacy Survey (DECaLS; Proposal ID 2014B-0404; PIs: David Schlegel and Arjun Dey), the Beijing-Arizona Sky Survey (BASS; NOAO Prop. ID 2015A-0801; PIs: Zhou Xu and Xiaohui Fan), and the Mayall z-band Legacy Survey (MzLS; Prop. ID 2016A-0453; PI: Arjun Dey). DECaLS, BASS and MzLS together include data obtained, respectively, at the Blanco telescope, Cerro Tololo Inter-American Observatory, NSF’s NOIRLab; the Bok telescope, Steward Observatory, University of Arizona; and the Mayall telescope, Kitt Peak National Observatory, NOIRLab. The Legacy Surveys project is honored to be permitted to conduct astronomical research on Iolkam Du’ag (Kitt Peak), a mountain with particular significance to the Tohono O’odham Nation.

This project used data obtained with the Dark Energy Camera (DECam), which was constructed by the Dark Energy Survey (DES) collaboration.

\facilities{\textit{WISE}, DESI, IRSA, \textit{Planck}, \textit{Spitzer}}

\software{astropy: \citet{2018AJ....156..123A}, CAMB: \citet{2000ApJ...538..473L}, colossus: \citet[][]{2018ApJS..239...35D}, Corrfunc: \citet{2020MNRAS.491.3022S}, emcee: \citet{2013PASP..125..306F}, healpy/HEALPix: \citet[][]{Zonca2019, 2005ApJ...622..759G}}, MANGLE: \citep[][]{2004MNRAS.349..115H, 2008MNRAS.387.1391S}, TOPCAT: \citep{2005ASPC..347...29T}.

\newpage


\bibliography{lensingQSOs}{}
\bibliographystyle{aasjournal}



\end{document}